\documentclass[]{aastex631}
\usepackage{verbatim}

\shorttitle{Data-Driven Simulation of NOAA AR 11283}
\shortauthors{Kang et al.}

\usepackage{amsmath} 
\begin{document}

\title{Data-driven MHD Simulation of the Formation of a Magnetic Flux Rope and an Inclined Solar Eruption}

\author[0009-0005-2172-0431]{Yeongmin Kang}
\affiliation{Institute for Space-Earth Environmental Research, Nagoya University,\\
Furo-cho, Chikusa-ku, Nagoya, Aichi 464-8601, Japan}

\author[0000-0002-7800-9262]{Takafumi Kaneko}
\affiliation{Niigata University, \\
8050 Ikarashi 2-no-cho, Nishi-ku, Niigata 950-2181, Japan}

\author[0000-0003-0026-931X]{K.D. Leka}
\affiliation{Institute for Space-Earth Environmental Research, Nagoya University,\\
Furo-cho, Chikusa-ku, Nagoya, Aichi 464-8601, Japan}
\affiliation{North West Research Associates\\
3380 Mitchell Lane, Boulder, CO 80301, USA}

\author[0000-0002-6814-6810]{Kanya Kusano}
\affiliation{Institute for Space-Earth Environmental Research, Nagoya University,\\
Furo-cho, Chikusa-ku, Nagoya, Aichi 464-8601, Japan}

\begin{abstract}

Solar energetic events are caused by the release of magnetic energy accumulated in the solar atmosphere. To understand their initiating physical mechanisms, the dynamics of the coronal magnetic fields must be studied. Unfortunately, the dominant mechanisms are still unclear due to lack of direct measurements.  Numerical simulations based on magnetohydrodynamics (MHD) can reproduce the dynamical evolution of solar coronal magnetic field providing a useful tool to explore flare initiation.  Data-driven MHD simulations, in which the time-series observational data of the photospheric magnetic field is used as the simulation boundary condition, can explore different mechanisms. To investigate the accumulation of free magnetic energy through to a solar eruption, we simulated the first of several large flares in NOAA Active Region 11283.  We used a data-driven model (\citealt{2021ApJ...909..155K}) that was governed by zero-beta MHD, focusing on the free magnetic energy accumulation prior to the M5.3 flare (September 6, 01:59 UT, 2011). We reproduced the flare-associated eruption following the formation of twisted magnetic fields, or a magnetic flux rope (MFR), formed by photospheric motions at its footpoints. We found that the eruption was first triggered by the growth of the torus instability.  The erupting MFR caused magnetic reconnections with neighboring magnetic field lines located along the direction of the eruption.  Using the simulation results and an axial-radial decay index centered on the MFR, we find a natural explanation for the inclination of the eruption and a possible approach to predict the direction of solar eruptive events. 

\end{abstract}


\newcommand{\fpp}[2]{\frac{\partial #1}{\partial #2}}
\newcommand{\vctr}[1]{\mbox{\boldmath $#1$}}

\newcommand*\subtxt[1]{_{\textnormal{#1}}}
\DeclareRobustCommand\_{\ifmmode\expandafter\subtxt\else\textunderscore\fi}

\section{Introduction} \label{sec:intro}

A solar flare is one of the energetic events on the Sun, exhibiting a sudden and rapid brightness increase across the electromagnetic spectrum. Based on magnetohydrodynamic (MHD) theories, the flare phenomenon is understood as the release of the magnetic energy stored in the solar atmosphere, and can accompany solar plasma eruptions \citep{2011LRSP....8....6S}. Numerous studies have been conducted to predict solar flares; for example, by finding a physics-based critical condition of the MHD instability \citep{2020Sci...369..587K} or by using empirical methods \citep[e.g.,][]{LekaBarnesWagner2018}. In order to improve flare predictions and gain a deeper understanding of the mechanisms of flare production, it is crucial to study the structure and dynamics of the coronal magnetic field in detail. However, the direct measurement of the coronal magnetic field remains challenging at present.

Several numerical methods have been developed to overcome this challenge. Nonlinear force-free field (NLFFF) modeling \citep[see review by][]{2016PEPS....3...19I} infers a coronal magnetic field assuming the ``force-free'' condition of $\vctr{J} \times \vctr{B}=0$, where $\vctr{J}$ and $\vctr{B}$ denote the electric current density and magnetic field vectors, respectively. Under this assumption, a three-dimensional coronal magnetic field is extrapolated from the photospheric vector magnetic field inferred by the observation. The mathematical requirement of $\vctr{J} \times \vctr{B}=0$ suggests that the electric current flows along the magnetic field lines. In the solar corona, the electric current is  generally considered to be a storage option for the free magnetic energy needed for solar eruptions. Thus, this approximation allows us to investigate the free magnetic energy and the magnetic helicity as a characteristic of the pre-eruptive phase. However, the force-free state may not be a proper assumption for ongoing solar eruptions since the photosphere drives active evolution of the coronal magnetic field, suggesting a non-force-free state in reality \citep{2018ApJ...865..146K}. 

A data-driven modeling approach (see review by \citealt{2022Innov...300236J}) uses time-series observational data as a temporally-varying boundary condition, allowing for a more realistic  evolution of the coronal magnetic field in response to changes in the photosphere. Several previous studies (e.g. \citealt{2012ApJ...757..147C}; \citealt{2015SpWea..13..369F}; \citealt{2016NatCo...711522J}; \citealt{2016ApJ...828...62J}; \citealt{2018ApJ...855...11H}; \citealt{2019ApJ...870L..21G}; \citealt{2020ApJS..250...28H}; \citealt{2021ApJ...909..155K}; \citealt{2022A&A...658A.200L}; \citealt{2023SPD....5410106A}) have employed different data-driven methods and data sets from various eruptive events. These studies have reproduced the dynamics of the three-dimensional coronal magnetic field. For example, \citet{2019ApJ...870L..21G} reproduced an eruption of sheared magnetic field lines, and their result agreed with the observation of their target event. \citet{2018ApJ...855...11H} succeeded in reproducing the coronal magnetic structure, however, their modeling could not reproduce the sudden release of magnetic energy (or the eruption), showing a disagreement with the observation. If results from data-driven simulations show a reasonable agreement with the observations, this modeling could offer a valuable approach to better understand the physical processes driving solar eruptions, including the dynamic evolution of magnetic structures during the eruption phases. Recently, \citet{2020ApJ...890..103T} conducted a comparative study of the results from different data-driven models employing different algorithms but sharing the same bottom boundary. They found significant differences in the outcomes, and this fact suggests both the challenges of data-driven modeling and the complexity of the solar physics. Therefore, continuing to improve the data-driven models is a meaningful work for understanding the physics of solar eruptions.

In the pre-flaring stages, changes of sunspot structures have been observed and numerically studied. For example, a rotating motion of photospheric magnetic field is a possible cause to produce flares (e.g. \citealt{2006JApA...27..277H}; \citealt{2013SoPh..286..453T}; \citealt{2016NatCo...713104L}; \citealt{2021ApJ...922..108J}). A magnetic flux cancellation is also another possible characteristics of the pre-flare stage (e.g. \citealt{2003PhPl...10.1971L}; \citealt{2012ApJ...760..101W}). In addition, an emergence of magnetic flux into the photosphere has been also proposed as a possible pre-flare event (e.g. \citealt{CanfieldPriestRust1975}; \citealt{1995JGR...100.3355F}; \citealt{2011LRSP....8....6S}; \citealt{2021RAA....21..312M}). All of the proposed process listed above can introduce new magnetic energy and complexity into a previously-existing magnetic field, potentially leading to the build-up of free magnetic energy and the development of unstable magnetic configurations. This accumulation of free magnetic energy, along with an increased non-potentiality of the magnetic field, can act as a driving force for flare production. For example, NOAA active region (AR) 11283 produced multiple large flares following the emergence of new magnetic flux. This region is of particular interest because the flux emergence transformed the photospheric magnetic field from a simple configuration to a more complex state before several significant flares occurred. Additionally, following the cessation of flux emergence, the flares occurred approximately 30 hours later. This delay indicates that a different process plays a role in energy accumulation. Figure \ref{fig:mag_of_m5.3} presents the normal component of photospheric vector magnetic fields obtained by the Helioseismic and Magnetic Imager \citep[HMI;][]{2012SoPh..275..229S} on board the Solar Dynamics Observatory \citep[SDO;][]{2012SoPh..275....3P} before and after the first major flare (M5.3 flare). Figure \ref{fig:obs_ar11283} shows the evolution of the photospheric magnetic field and the observational images in the different wavelength pass-bands obtained by the Atmospheric Imaging Assembly \citep[AIA;][]{2012SoPh..275...17L} on board the SDO, displaying the same field of view (FOV). Figure \ref{fig:goes} shows the peak intensity of the M5.3 flare in the AR, observed by the Geostationary Operational Environmental Satellites (GOES) and the temporal evolution of magnetic flux in the FOV of Figure \ref{fig:mag_of_m5.3} (b). As depicted in Figure \ref{fig:goes}, the initial significant flare occurred at 01:48 UT on September 6, approximately 30 hours after the termination of the flux emergence. We will provide a detailed description of the observational features in Section \ref{sec:event}.

Previous studies (\citealt{2013ApJ...771L..30J}; \citealt{2016NatCo...711522J}; \citealt{2020ApJ...903..129P}) have conducted numerical simulations of this AR, focusing on the large flares including the M5.3 and X2.1 flares on September 6, 2011. In particular, \citet{2016NatCo...711522J} performed a data-driven simulation of the M5.3 flare, which was followed by a filament ejection approximately 1 hour later. They solved a full set of time-dependent MHD equations and used a NLFFF model extrapolated from the photospheric magnetic field data 50 hours before the flare as the initial condition. Their simulation reproduced not only a bundle of twisted magnetic field, referred to as a magnetic flux rope (MFR), over the ``polarity inversion line'' (PIL) between the positive and negative magnetic poles (see Figure \ref{fig:obs_ar11283}, a) but also the following flare-related eruption. An MFR has been suggested as being an important structure associated with solar eruptions (\citealt{2006SSRv..123..251F}; \citealt{2006ApJ...641..577M}; \citealt{2008A&A...492L..35A}; \citealt{2019LRSP...16....3T}). In \citet{2016NatCo...711522J}'s modeling, however, the flare reconnection occurred between overlying fields situated above the MFR and other field lines located on the north-west side of the MFR, with the reconnection site located in the western part of the AR. In other words, they proposed that the M5.3 flare-related eruption was not triggered solely by the loss of stability in the magnetic field of the MFR but triggered by some reconnections which happened between the MFR's field lines and other neighbor field lines. However, what mechanism was most important and the causality may still remain a question because the first flare-manifest brightening in the chromosphere, exhibiting a typical structure of flare ribbons, appeared along the PIL at September 6, 01:48 UT, 2011, close to the flare peak time (see Figure \ref{fig:obs_ar11283}, b). In other words, the observed location of the flare ribbons do not agree with the footpoints of the reconnected field lines in their simulation, which is where chromospheric brightening would be expected. As such, deduced from the location of the flare ribbons, it is still possible that the twisted field lines above the PIL, inferred from the photospheric magnetic field, should play a role in the flare initiation. For example, \citet{2023SPD....5410106A} performed a data-driven simulation of the flare event and has found that the evolution of their simulated magnetic field showed an agreement with the evolution of the observed flare ribbons. Therefore, revisiting this event by investigating alternative mechanisms of flare-related eruptions is a valuable contribution to understanding the causal relationship between flare reconnection and eruptions.

There are two primary objectives in this study. The first objective is to investigate the accumulation of free magnetic energy prior to, and the initiation mechanism of, the M5.3 flare in AR 11283. To this end, we conducted a long-term data-driven MHD simulation using time-series of observational magnetic field data. The simulation period covered 36 hours leading up to and including the M5.3 flare. In the simulation, we have reproduced the formation of sheared magnetic field structures above the PIL and an eruptive event corresponding to the M5.3 flare. The second objective is to further develop a new method to explain and predict the direction of eruptions through the data-driven modeling. Further, we introduce a new methodology for calculating the decay index (\citealt{1978mit..book.....B}) that considers all possible directions of the eruption, allowing a prediction of the direction of eruption.

In Section \ref{sec:event}, we present the observational properties of the target AR and the evolution of its photospheric magnetic structure from 30 hours before the M5.3 flare until its onset. Section \ref{sec:model} describes the simulation method, followed by the presentation of numerical results in Section \ref{sec:results}. In Section \ref{sec:discussions}, we analyse our results using the new decay index calculation and discuss our modeling in comparison to the previous study. In Section \ref{sec:summary}, a summary of this study is provided.

\section{event} \label{sec:event}

\begin{figure}[htbp!] 
\centering    
\includegraphics[width=1.00\textwidth]{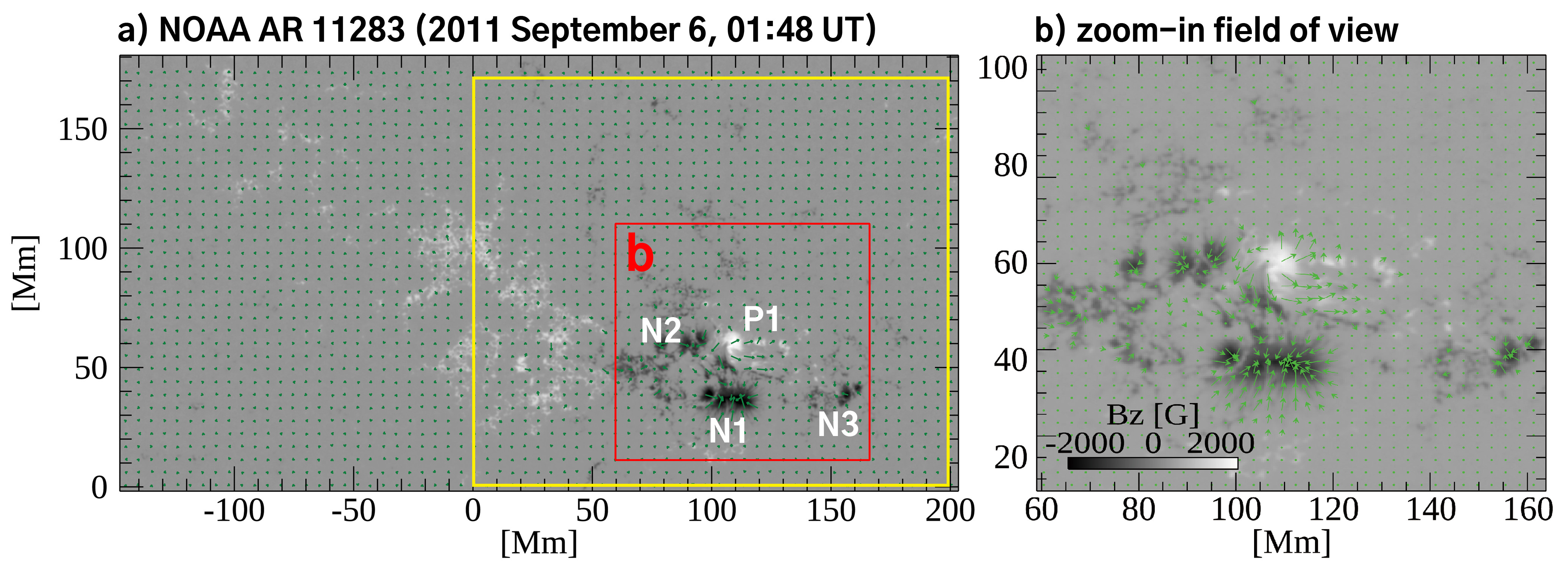}
\caption{Photospheric vector magnetic field of the M5.3 flaring region in NOAA AR 11283 on September 6, 01:48 UT, 2011 observed by SDO/HMI (from the {\tt hmi.sharp\_cea\_720s} data series). The gray scale represents the vertical component of photospheric magnetic field and the green arrows indicate the horizontal magnetic field; the minimum and maximum magnitudes of the horizontal fields shown by the arrows are 300 G and 1500 G, respectively. Panel (a) shows the full HARP area. The yellow-boxed region in panel (a) displays the simulation domain. Panel (b) shows the red-boxed region in panel (a). P1 corresponds to the location of a group of positive magnetic field concentrations in the center of image. N1, N2 and N3 correspond to the location of three negative concentrations around P1, with N1 to the south, N2 located to the north, and N3 located to the south-west of P1.}
\label{fig:mag_of_m5.3}
\end{figure}

\begin{figure}[htbp!] 
\centering    
\includegraphics[width=1.00\textwidth]{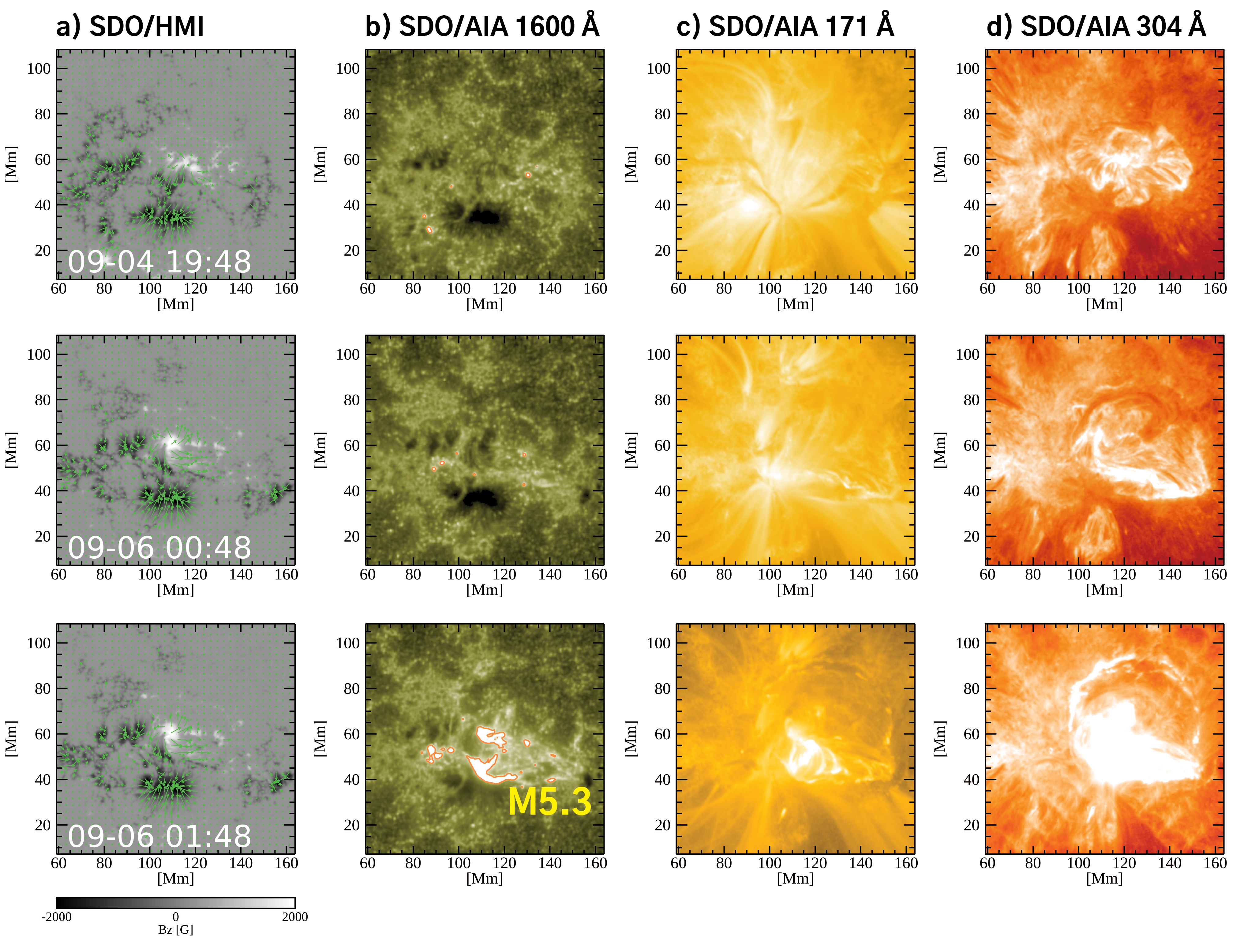}
\caption{NOAA AR 11283 and its observed temporal evolution. Panel (a) shows the SDO/HMI vector magnetograms, following Figure \ref{fig:mag_of_m5.3}. The background image shows the vertical magnetic field. Panels (b), (c) and (d) show SDO/AIA 1600 {\AA}, 171 {\AA} and 304 {\AA} images, respectively. The SDO/AIA images are displayed in a logarithmic scale and have the same FOV of panel (a), and have been coaligned to the magnetograms in CEA projection (neglecting their 3-D projection for this near-disk-center region). In panel (b), the bright flare ribbons, indicated by the red contours, were observed along the PIL on September 6, 01:48 UT, 2011, close to the flare peak time. Movie 1 displays the temporal evolution of all the panels in the entire period from September 4, 19:48 UT, 2011 to September 6, 06:48 UT, 2011.}

\label{fig:obs_ar11283}
\end{figure}

\begin{figure}[htbp!] 
\centering    
\includegraphics[width=0.80\textwidth]{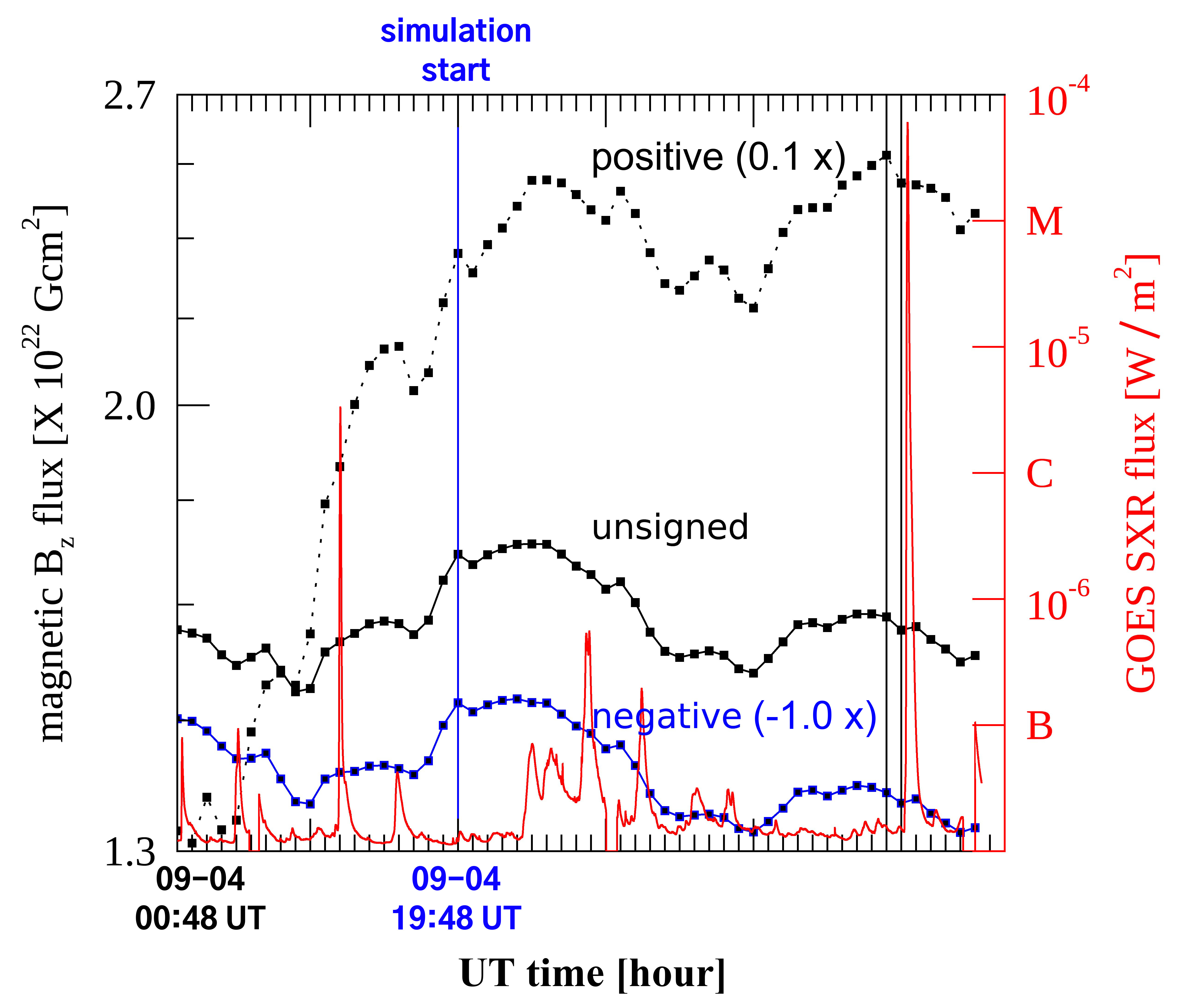}
\caption{X-ray light curves observed by the GOES XRS instrument in the 1--8\AA\ band, from September 4, 00:48 UT, 2011 to September 6, 06:48 UT, 2011 with the temporal evolution of magnetic flux on the photosphere. The data is plotted at 1 hour intervals. The horizontal axis is the observational time. Our simulation starts at September 4, 19:48 UT, 2011. The left-hand (right-hand) vertical axes indicate the magnetic flux (the GOES flux intensity), respectively. The solid (dashed) black lines are the temporal evolution of unsigned magnetic flux (positive magnetic flux) in the FOV of Figure \ref{fig:mag_of_m5.3} (b). The solid blue line is the temporal evolution of negative magnetic flux in the same FOV. The solid red line is the GOES X-ray light curve. The vertical lines sequentially correspond to the observational times of the photospheric magnetic field images shown in Figure \ref{fig:obs_ar11283} (a). Among the vertical lines, the blue one corresponds to the time of initial condition of simulation. The M5.3 flare which occurred in AR\,11283 had a peak of flux intensity at around September 6, 01:59 UT, 2011. No larger flares than the class-C were observed in the simulation period.}
\label{fig:goes}
\end{figure}

NOAA AR 11283, our simulation target, exhibited several large flares between September 6, 2011 and September 7, 2011. The first significant flare produced a GOES Soft X-ray peak intensity of $5.3\times10^{-5}\,{\rm W\,m}^{-2}$ on September 6, 01:59 UT, 2011 (see Figure \ref{fig:goes}). Figure \ref{fig:obs_ar11283} (b) shows an image from SDO/AIA in the 1600 {\AA} filter showing emission from the chromospheric or low coronal region. Notably, bright flare ribbons were produced along the PIL starting at approximately September 6, 01:48 UT, 2011, a time close to the peak of the M5.3 flare. We defined the flare ribbons as areas having a brightness greater than $1.4 Q(t)$ where $Q(t)$ is the median value of all pixels in the 1600 {\AA} pass-band within the FOV of the yellow-boxed region indicated in Figure \ref{fig:mag_of_m5.3} (a) at an observational time $t$. The similar method for defining flare ribbons has also been employed in \citet{2017ApJ...845...49K}. The SDO/AIA 171 {\AA} and 304 {\AA} images, shown in Figure \ref{fig:obs_ar11283} (c) and Figure \ref{fig:obs_ar11283} (d), respectively, sampling the upper corona, and show a corresponding brightness increase within the same FOV.

Considering the location of the flare ribbons (Figure \ref{fig:obs_ar11283}, b), we infer that four magnetic regions (each assumed to be comprised of unresolved groups of magnetic concentrations), denoted P1, N1, N2 and N3 (Figure \ref{fig:mag_of_m5.3}, a), were involved with the flare, and prior to the flare, P1 moved eastward. Surrounding P1, there were three negative concentrations (N1, N2 and N3) situated on the east, south, and southwest sided of P1, respectively. Notably, the convergence of P1 towards N1, driven by photospheric motions, resulted in a change in the shape of the PIL formed between P1 and N1, injecting the magnetic helicity in the local region. Inferred from Movie 1, the M5.3 flare was produced subsequent to this convergence, suggesting that the production of the M5.3 flare is associated with this converging motion which may have yielded  sheared field lines located above the PIL.

According to the Mount Wilson classification scheme \citep{1919ApJ....49..153H}, this active region was initially classified as a $\beta$-type region but changed to a $\beta\gamma\delta$-type region following the flux emergence near P1. It is well-known that $\delta$-type regions have more complex magnetic field configurations and tend to produce larger flares compared to regions with simpler magnetic distributions \citep{1987SoPh..113..267Z,2000ApJ...540..583S}. The flux emergence could form twisted field lines above the PIL (\citealt{1996ApJ...462..547L}; \citealt{2008ApJ...673L.215O}), which can increase the active region's non-potentiality. Moreover, the observation showed the eastward advection of P1 prior to the major flare occurrences. These observational features suggest that the evolution of magnetic field near P1 could be related to the production of following large flares. Through our data-driven modeling and the analysis of simulation results, we investigate the cause of the change in the magnetic field in this portion and its relationship with the following flare initiation.

\section{model} \label{sec:model}
The numerical model is based on the $\vctr{E} \times \vctr{B}$ driven method \citep{2021ApJ...909..155K}. In this model, an electric field ($\vctr{E}$) is determined by solving the Faraday's law inversely \citep{2010ApJ...715..242F} using time-series photospheric magnetic field maps (full magnetic field vector) as the bottom boundary. An inverted velocity field ($\vctr{v^{I}}$) is then computed using the equation $\vctr{v^{I}}=(\vctr{E} \times \vctr{B})/B^{2}$ in the bottom boundary of the simulation domain. In this study, we solved the MHD equations based on the zero-$\beta$ model under the assumption of the time-invariant plasma density as follows:

\begin{equation}
    \fpp{(\rho\vctr{v})}{t}+\nabla\cdot\left(\rho\vctr{v}\vctr{v}+\frac{B^{2}}{8\pi}\vctr{I}-\frac{\vctr{B}\vctr{B}}{4\pi}\right)=0,
    \label{eq:momentum}
\end{equation}

\begin{equation}
    \fpp{\vctr{B}}{t}=\nabla\times(\vctr{v}\times\vctr{B}-\eta\vctr{J}),
    \label{eq:induction}
\end{equation}

\begin{equation}
    \vctr{J}=\frac{1}{4\pi}\nabla\times\vctr{B},
    \label{eq:current}
\end{equation}

 \noindent where $\rho$, $t$, $\vctr{v}$, $\vctr{B}$, $\eta$, $\vctr{J}$ and $\vctr{I}$ represent mass density, time, a velocity field, a magnetic field, a resistivity, a current density, and an unit vector, respectively. We also applied an anomalous resistivity as follows:
 
\begin{equation}
  \begin{split}
    \eta=0 \quad \mathrm{for} \quad J < J_{\mathrm{th}}, \\
    \eta=\eta_{0}(J/J_{\mathrm{th}}-1)^{2} \quad \mathrm{for} \quad J\geq J_{\mathrm{th}},  
  \end{split}
  \label{eq:anomalous}
\end{equation}

 \noindent where $J_{\mathrm{th}}=10^{-8}$ $\mathrm{G/cm}$ and $\eta_{0}=10^{12}$ $\mathrm{cm^{2}/s}$, and we set the maximum value of the resistivity as $10^{12}$ $\mathrm{cm^{2}/s}$. In addition, we applied the divergence cleaning method \citep{2002JCoPh.175..645D} to suppress the artifact due to the deviation from the solenoidal condition $\nabla\cdot\vctr{B}=0$.
 
 The simulation domain was constructed in a three-dimensional Cartesian coordinate system in which the $xy$-plane is a horizontal plane parallel to the photosphere, and the $z$-direction corresponds to the height from the photosphere.  The computational box extends $0<x<205.2$ Mm, $0<y<176.4$ Mm, $-1.44$ Mm $<z<64.8$ Mm, uniformly resolved by $N_{x}\times N_{y}\times N_{z}=570\times490\times190$ grid points, thus uniform grid spacing size of 0.36 Mm in all directions. We adopted a periodic boundary condition to the $x$ and $y$-directions and a free boundary condition for the top boundary. We mimicked a stratified density profile as $\rho=\rho_{0}\exp[-z/H]$, where $\rho_{0}=1.67\times10^{-15}$ $\mathrm{g/cm^{3}}$ and the scale height $H=30$ Mm.

 This modeling used the HMI Active Region Patches \citep[HARP;][]{2014SoPh..289.3549B} magnetic field cutouts\footnote{series {\tt hmi.sharp\_cea\_720s}, see {\tt http://jsoc.stanford.edu/ajax/exportdata.html}} (see Figure \ref{fig:mag_of_m5.3}) as the input data. However, shown in the yellow-boxed region in Figure \ref{fig:mag_of_m5.3} (a), our simulation box was smaller than the HARP because we focused on the magnetic fields associated with the twisted field lines above the PIL so that the western part where \citet{2016NatCo...711522J} discussed as the region of reconnection related to the subsequent eruption was excluded in this paper. To minimize the data noise and numerical errors, we applied a low-pass filter with a cut-off length of $\lambda>4.41$ Mm corresponding to Fourier mode 40 in the $y$-direction. We also applied another filter on the inverted electric field as follows:

\begin{gather}
\begin{aligned}
&\begin{gathered}
\vctr{E^{f}}=\vctr{E^{I}}f(B), \\
f(B)=0.5\left[1+\tanh\left(\frac{B-B_{\mathrm{th}}}{w}\right)\right],
\end{gathered}
\end{aligned}
\label{eq:inversion_filter}
\end{gather}

\noindent
 where $\vctr{E^{f}}$, $\vctr{E^{I}}$ and $f(B)$ represent the filtered electric field, the inverted electric field, and the filter function depending on the field strength $|\vctr{B}|$, respectively. We set $B_{\mathrm{th}}=200$ G and $w=200$ G. Thus, we decreased the strength of the electric field where $|\vctr{B}|<200$ G, minimizing the influence of data noise and consequently reducing unrealistic velocity field reproductions. Additionally, we reduced the magnetic field strength of the low-pass filtered data by a factor of 50 to make the computation faster. This method was similarly employed in \citet{2016NatCo...711522J}. The original time cadence of HMI pipeline data is 720 seconds, but we only sampled the time-series data at a lower cadence of 1 hour. The simulation period was from September 4, 19:48 UT, 2011 to September 6, 06:48 UT, 2011 covering both the pre-flaring stage and the M5.3 flare itself. Moreover, for the numerical efficiency, we set the time interval between two magnetic maps to be 360 seconds in the simulation time unit, or 10 times shorter than the reality. Therefore, the photospheric magnetic field in the simulation evolves 10 times faster than in reality, however, the photospheric velocity in the simulation is still much slower than the Alfvén speed in the corona, and the changes in the photosphere immediately propagate to the coronal region as they do in reality.

\begin{figure}
      
\centerline{
\includegraphics[width=0.50\textwidth,clip,trim=50mm 20mm 0mm 20mm, angle=0]{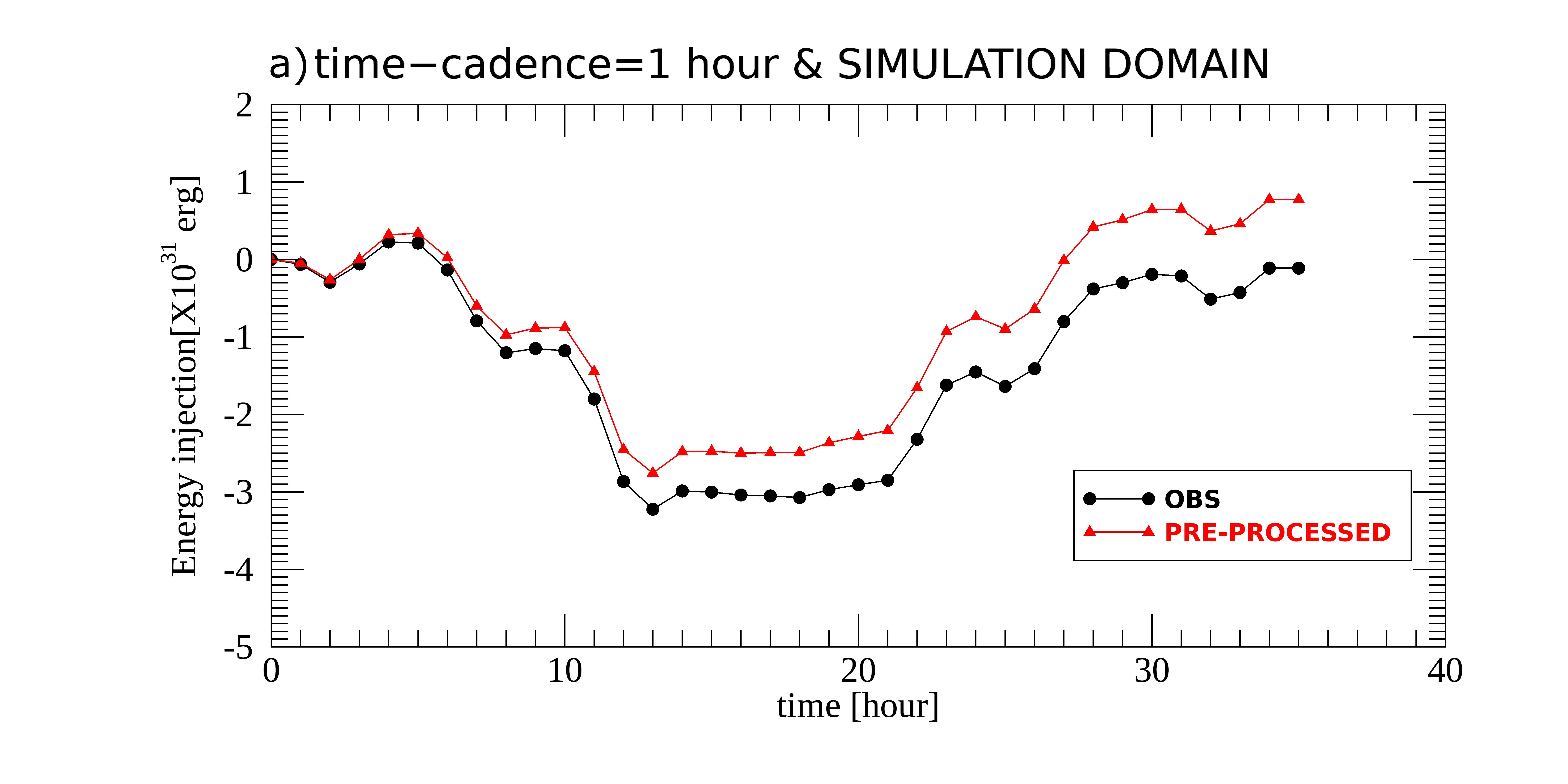}
\includegraphics[width=0.50\textwidth,clip,trim=50mm 20mm 0mm 20mm, angle=0]{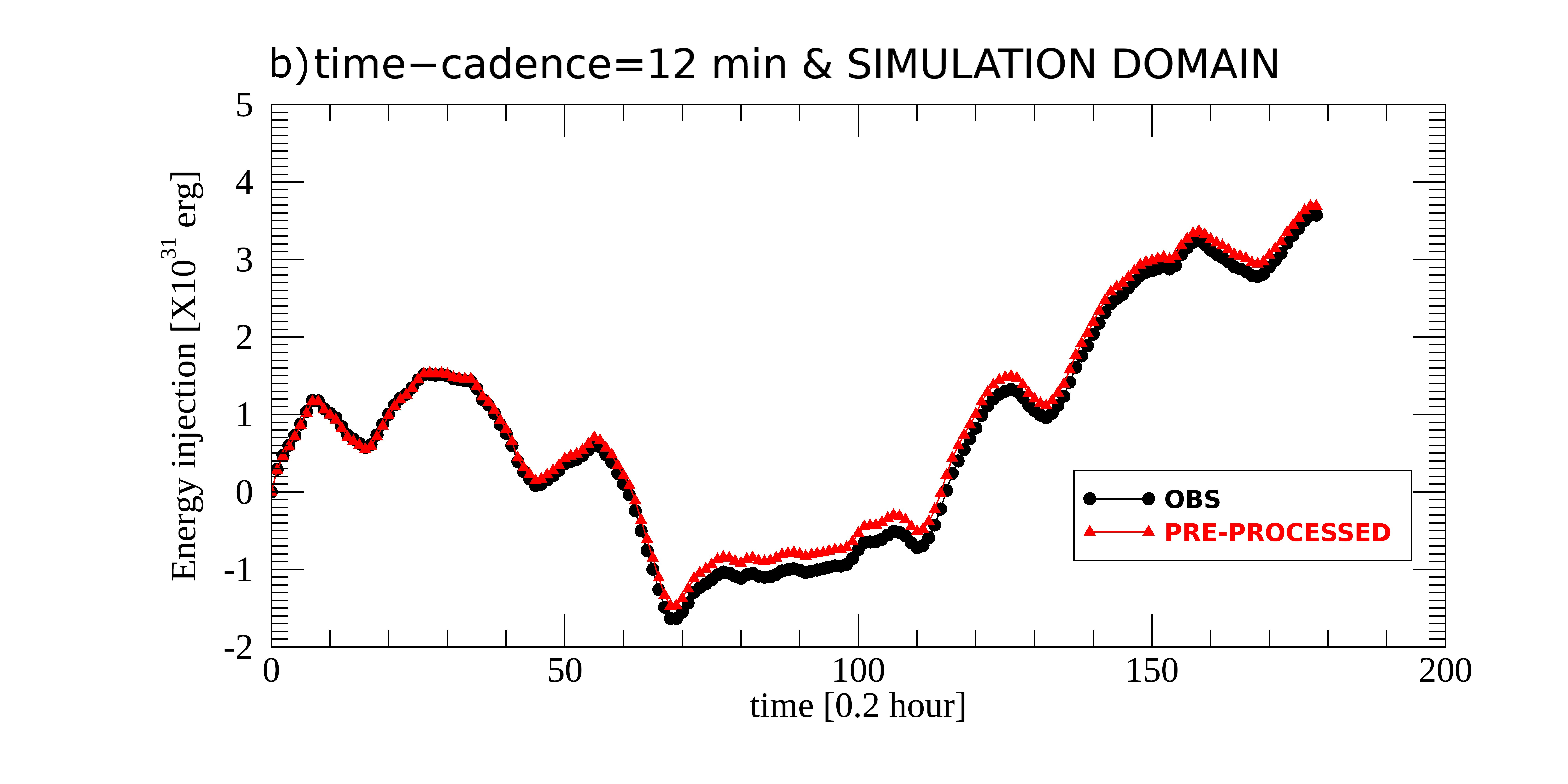}}
\centerline{
\includegraphics[width=0.50\textwidth,clip,trim=50mm 20mm 0mm 20mm, angle=0]{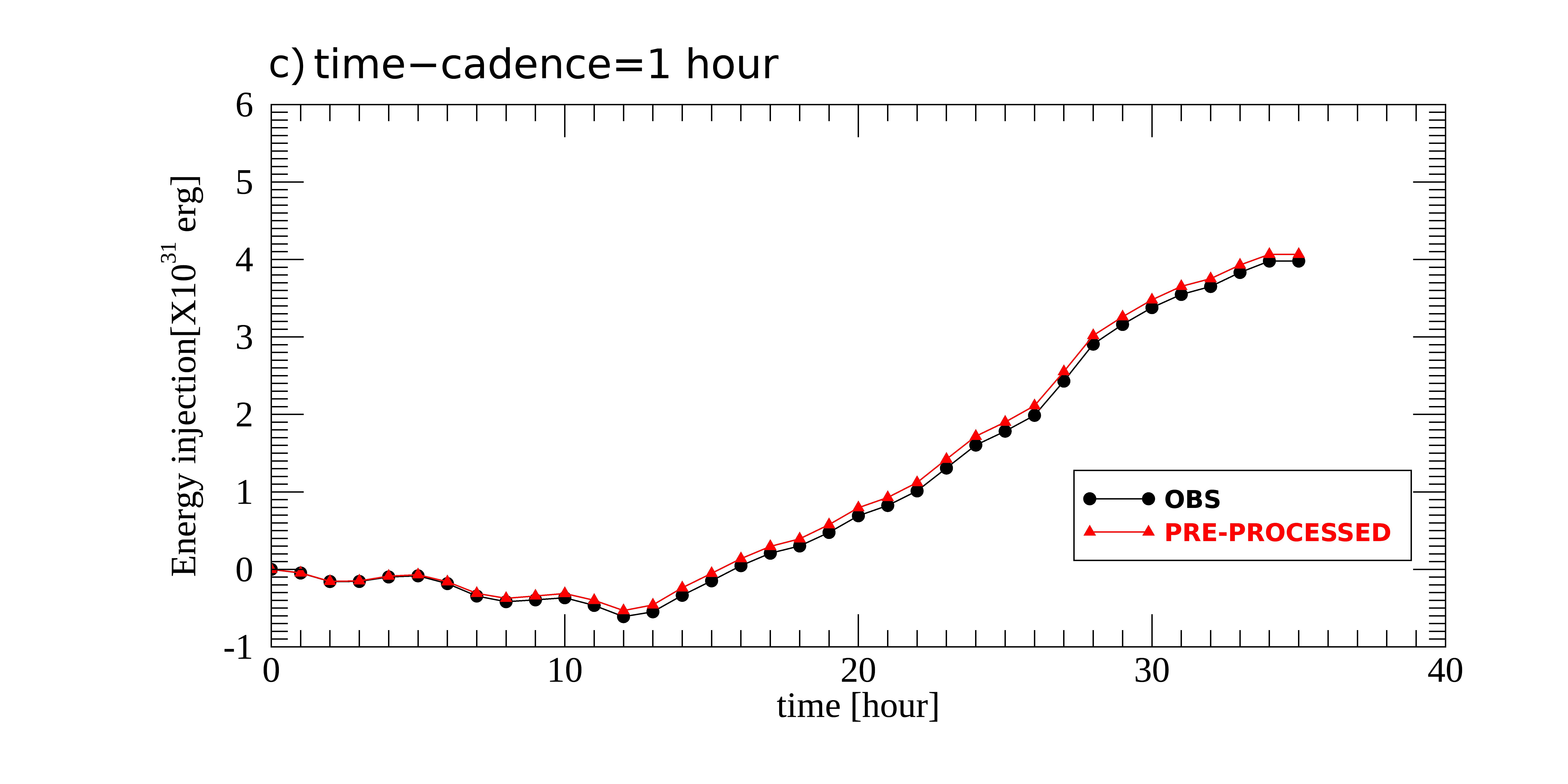}
\includegraphics[width=0.50\textwidth,clip,trim=50mm 20mm 0mm 20mm, angle=0]{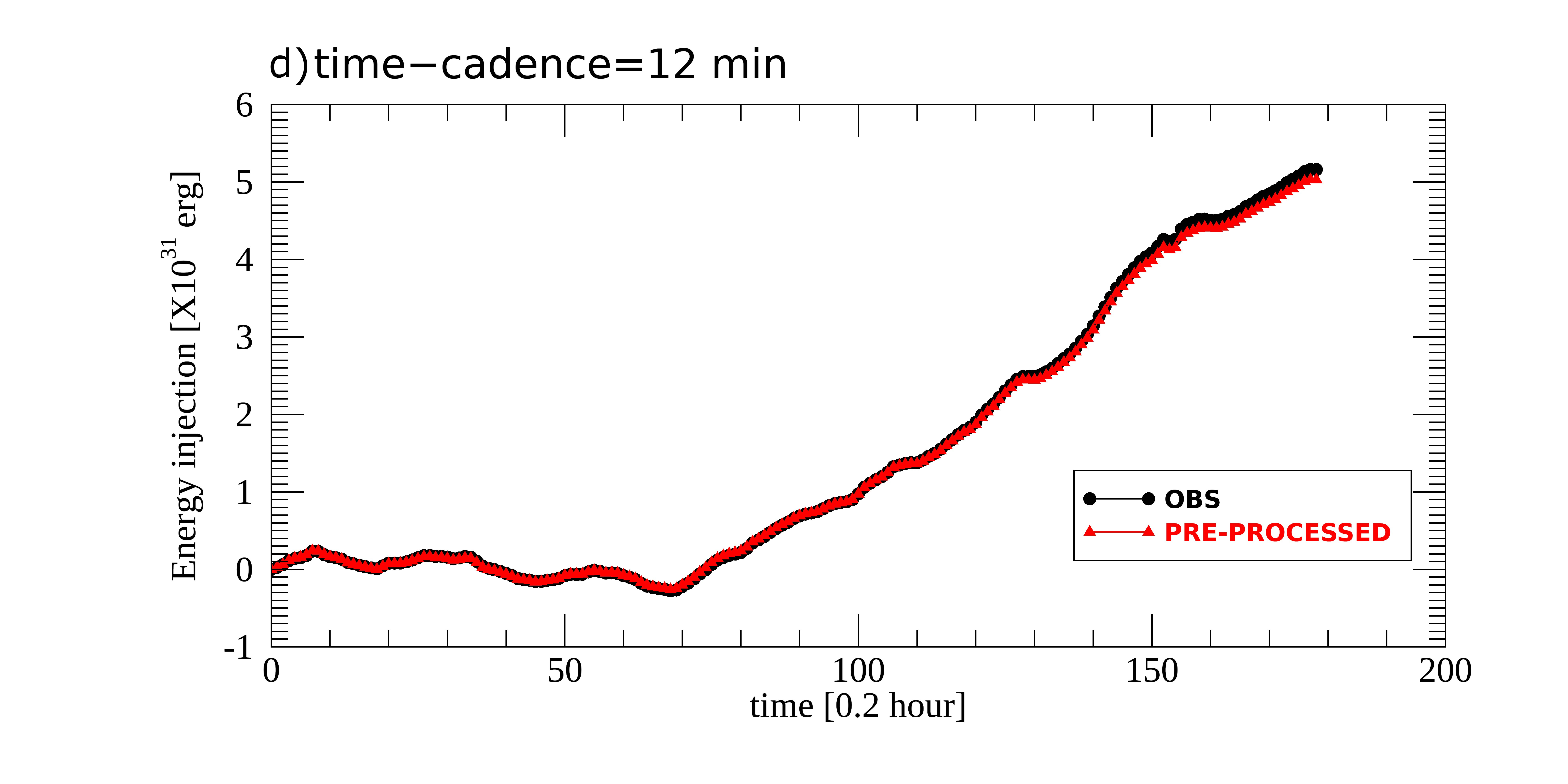}}
                
\caption{The temporal evolution of energy injection. The black (red) solid lines indicate the temporal evolution of energy injection in the observation (pre-processed data). Time t=0 marks the start of the simulation, corresponding to September, 19:48 UT, 2011 in the observational time. (a) The case of 1 hour cadence within the simulation domain. (b) The case of 12 minutes cadence within the simulation domain. (c) The case of 1 hour cadence within the red-boxed region in Figure \ref{fig:bznp} where the HiFER is focused (the FOV of Figure \ref{fig:bznp_mag}). (d) The case of 12 minutes cadence within the red-boxed region in Figure \ref{fig:bznp} (the FOV of Figure \ref{fig:bznp_mag}).}
\label{fig:downsampling_pre-processed}    
\end{figure}

 In \citet{2019SoPh..294...84L} it was suggested that using a different time cadence of observed magnetic field data to calculate a velocity field can cause a variation of energy injection rate, and using pre-processed data as the lower bottom boundary may also affect the energy injection. To address this concern, we compared the temporal evolution of injected energy for two datasets, the HARP data and the pre-processed (the low-pass filtered) data  which have the different time cadences, 12 minutes and 1 hour, respectively. The photospheric energy injection is given by as follows:

 \begin{equation}
    E\_{injection}=\int_{0}^{t} \int \vctr{S} \cdot \vctr{\hat{z}}dAdt' = \int_{0}^{t} \int \frac{c}{4\pi} (\vctr{E} \times \vctr{B}) \cdot \vctr{\hat{z}}dAdt'
    \label{eq:energy injection}
\end{equation}

\noindent
 where $c$, $\vctr{S}$ and $\vctr{\hat{z}}$ denote the speed of light, the Poynting vector on the photosphere and the unit vector along the $z$-direction, respectively. The normal component of Poynting vector can be given by $S_{n}=(c/4\pi)(-(\vctr{v_{h}} \cdot \vctr{B_{h}})B_{n}+B_{h}^{2}v_{n})$ where the subscripts $n$ and $h$ denote the normal component and the horizontal components with respect to the photosphere, respectively. The velocity field used here was obtained by our data-driven algorithm for computing the velocity field ($\vctr{v^{I}}$). The surface integration ($\int dA$) was conducted within the red-boxed region indicated in Figure \ref{fig:bznp}, which encompasses the high free magnetic energy region \citep[HiFER;][]{2020Sci...369..587K} where the non-potential component of the magnetic field is greater than a certain threshold (see Figure \ref{fig:bznp}, Figure \ref{fig:bznp_mag} and Section \ref{sec:results} for the location of integral area and the detailed definition of HiFER). Figure \ref{fig:downsampling_pre-processed} presents the comparison of energy injection evolution for the different time cadences in the cases of both the raw and pre-processed data. Figure \ref{fig:downsampling_pre-processed} (a) and (b) depict the energy injection across the entire simulation domain at intervals of 1 hour and 12 minutes, respectively. Figure \ref{fig:downsampling_pre-processed} (c) and (d) show the energy injection within the HiFER-focused region at intervals of 1 hour and 12 minutes, respectively. Inferred from Figure \ref{fig:downsampling_pre-processed} (a) and (b), the injected energy at a 12-minutes cadence was slightly greater than the case of 1-hour cadence, but both showed similar overall trends. When comparing the observed and pre-processed data across all panels, it is evident that the pre-processing did not display significant differences across different temporal resolutions. Additionally, comparing Figure \ref{fig:downsampling_pre-processed} (a) and (c), or Figure \ref{fig:downsampling_pre-processed} (b) and (d), the different trends shown between the two investigated areas indicate that while the energy injection in the entire simulation domain dropped below zero most of period, it consistently remained above zero within the HiFER-focused region from t=15 hr onwards. This distinct temporal evolution of energy injection suggests that the increase of energy injection in the simulation domain from t=12 hr was mainly driven by the HiFER-focused region. In conclusion, we confirmed that only the downsampling mainly affected the quantity of injected energy into the HiFER, but the trend of energy injection did not show the significant change. This suggests that the simulation will be able to reproduce the expected energy injection.

 Before conducting the main simulation in which the magnetic fields were reproduced in the simulation domain, we carried out a pre-simulation to convert the potential field to the magnetic field in the first snapshot. In the pre-simulation stage, at the bottom boundary, a potential field is gradually changed to reproduce the horizontal field of the first observational snapshot. The potential field was calculated by Fourier expansion method \citep{2014masu.book.....P} only using the vertical component of the photospheric magnetic field. We used a four-step Runge-Kutta method \citep{2005A&A...429..335V} and a fourth order central finite difference scheme to solve the MHD equations numerically, and an artificial viscosity \citep{2014ApJ...789..132R} was adopted to suppress the numerical oscillation.

\newpage
\section{results} \label{sec:results}

\begin{figure}[htbp!] 
\centering    
\includegraphics[width=0.80\textwidth]{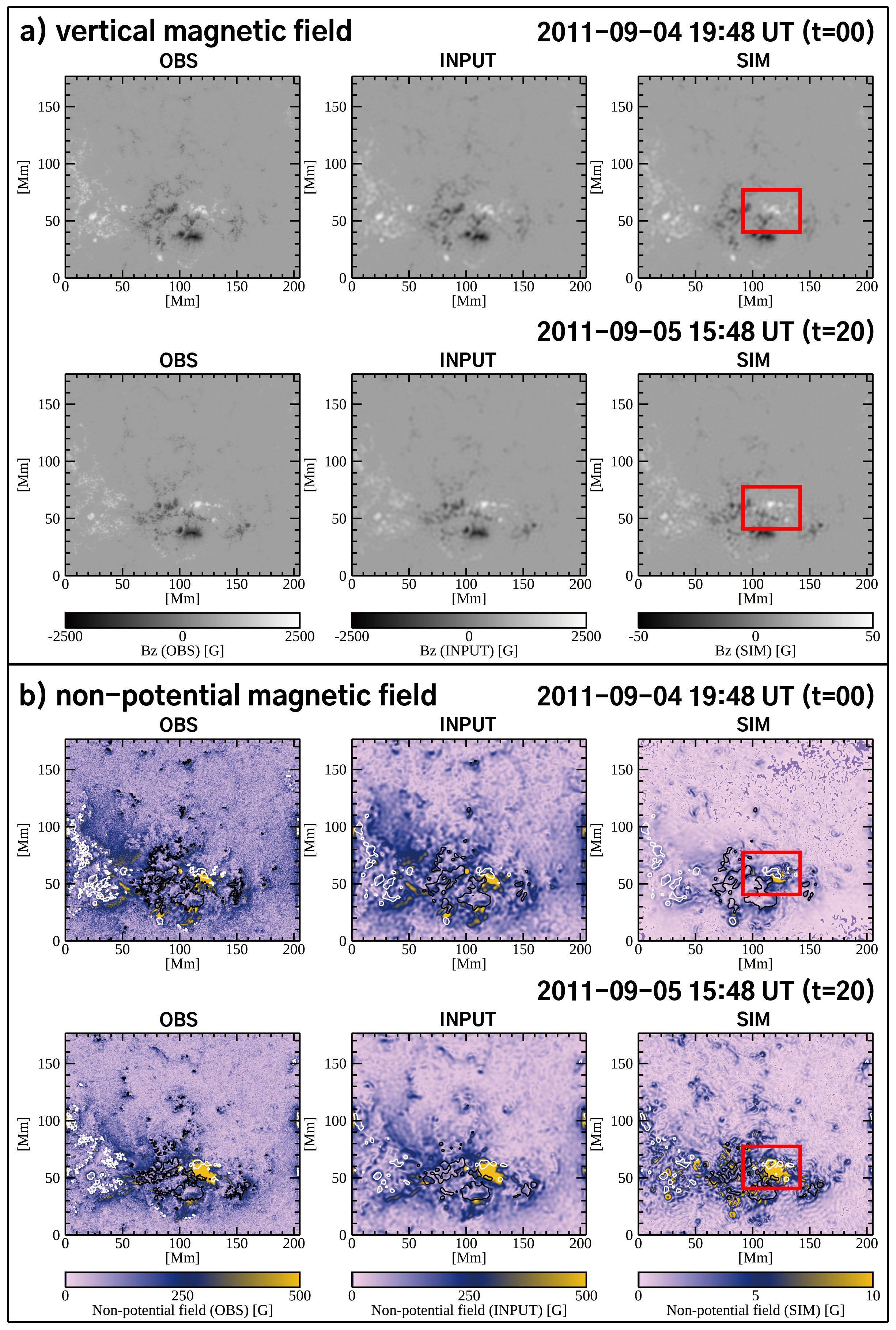}
\caption{(a) The photospheric magnetic fields ($B_{z}$) at t=0 and 20 hr. (b) The photospheric $B_{\mathrm{np}}$ at t=0 and 20 hr. The observational data (left), the input (middle) and the simulation result (right) are shown in both panels (a) and (b). The red boxes indicate the regions that encompass HiFER. Note that the color scales of the simulated magnetic field and $B_{\mathrm{np}}$ are reduced by a factor of 50. The white (black) contours in panel (b) indicates the $B_{z}$ of 500 G (-500 G) for the observation and the input and 10 G (-10 G) for the simulation result.}
\label{fig:bznp}
\end{figure}

Our modeling has reproduced both the temporal evolution of photospheric magnetic field and the accumulation of free magnetic energy on the photosphere. Figure \ref{fig:bznp} shows a comparison of the simulation results to both the observation and the input to the simulation. Note that the input refers to the magnetic field used for the velocity inversion (the smoothed observational data). Figure \ref{fig:bznp} (a) shows the temporal evolution of the photospheric $B_{z}$ and Figure \ref{fig:bznp} (b) shows the photospheric non-potential magnetic field ($B_{\mathrm{np}}$). The non-potential magnetic field, which shows the non-potentiality of the photospheric magnetic field, is given by $B_{\mathrm{np}}=\sqrt{(B_{x}-B_{\mathrm{pot,}x})^{2}+(B_{y}-B_{\mathrm{pot,}y})^{2}}$ where $B_{\mathrm{pot}}$ indicates the potential magnetic field. Thus, the accumulation process of free magnetic energy can be described by the temporal evolution of $B_{\mathrm{np}}$.

\begin{figure}[htbp!] 
\centering    
\includegraphics[width=1.00\textwidth]{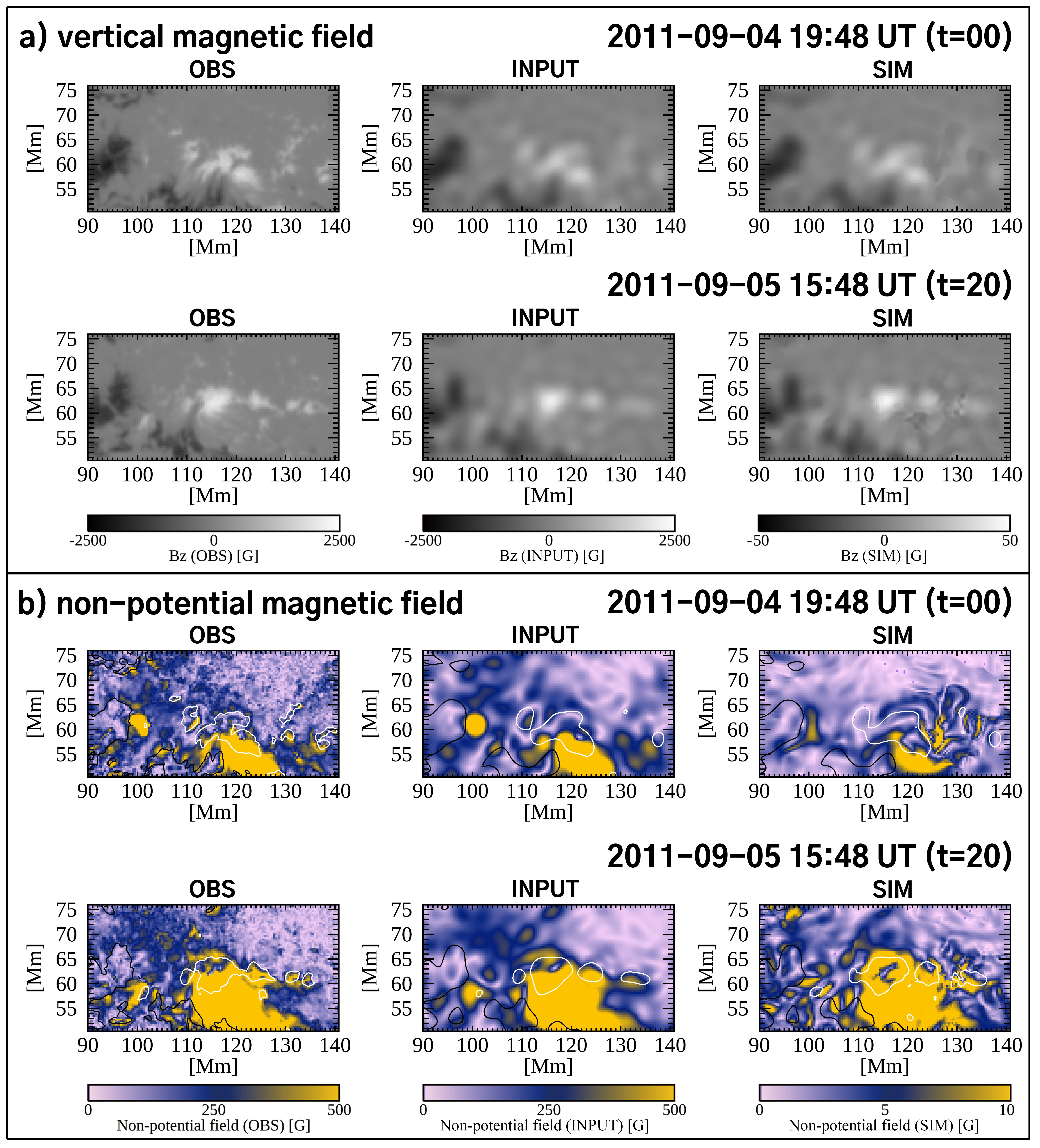}
\caption{Zoom-in FOV of the red-boxed region indicated in Figure \ref{fig:bznp}. (a) The photospheric magnetic fields ($B_{z}$) at t=0 and 20 hr. (b) The photospheric $B_{\mathrm{np}}$ at t=0 and 20 hr. The observational data (left), the input (middle) and the simulation result (right) are shown in both panels (a) and (b). Same as Figure \ref{fig:bznp}, the white (black) contours in panel (b) indicates the $B_{z}$ of 500 G (-500 G) for the observation and the input and 10 G (-10 G) for the simulation result.}
\label{fig:bznp_mag}
\end{figure}

In Figure \ref{fig:bznp} (b), one can see that the non-potentiality of the magnetic field concentrates both near the PIL formed by P1 and N2, and over P1 itself. Consistent with the inferences from the SDO/HMI and SDO/AIA 1600 {\AA} observations, the result suggests that the M5.3 flare production was related to the HiFER defined by the areas exhibiting $B_{\mathrm{np}}>8$ G in the simulation results (Figure \ref{fig:bznp}, b), corresponding to regions where the non-potential field intensity is greater than 40 G in reality. The red boxes in Figure \ref{fig:bznp} are areas that encompass these pixels designated as HiFER, set at $90 \; \mathrm{Mm} <x<147.6$ Mm and $39.6 \; \mathrm{Mm} <y<82.8$ Mm on the simulated photosphere. We fixed both the location and the size of red-boxed region for the full time-series because this region fully covered the evolution of the flare-related $B_{\mathrm{np}}$ over the whole simulation period. Figure \ref{fig:bznp_mag} displays the zoom-in FOV of the red-boxed regions designated in Figure \ref{fig:bznp}.

\begin{figure}[htbp!] 
\centering    
\includegraphics[width=1.00\textwidth]{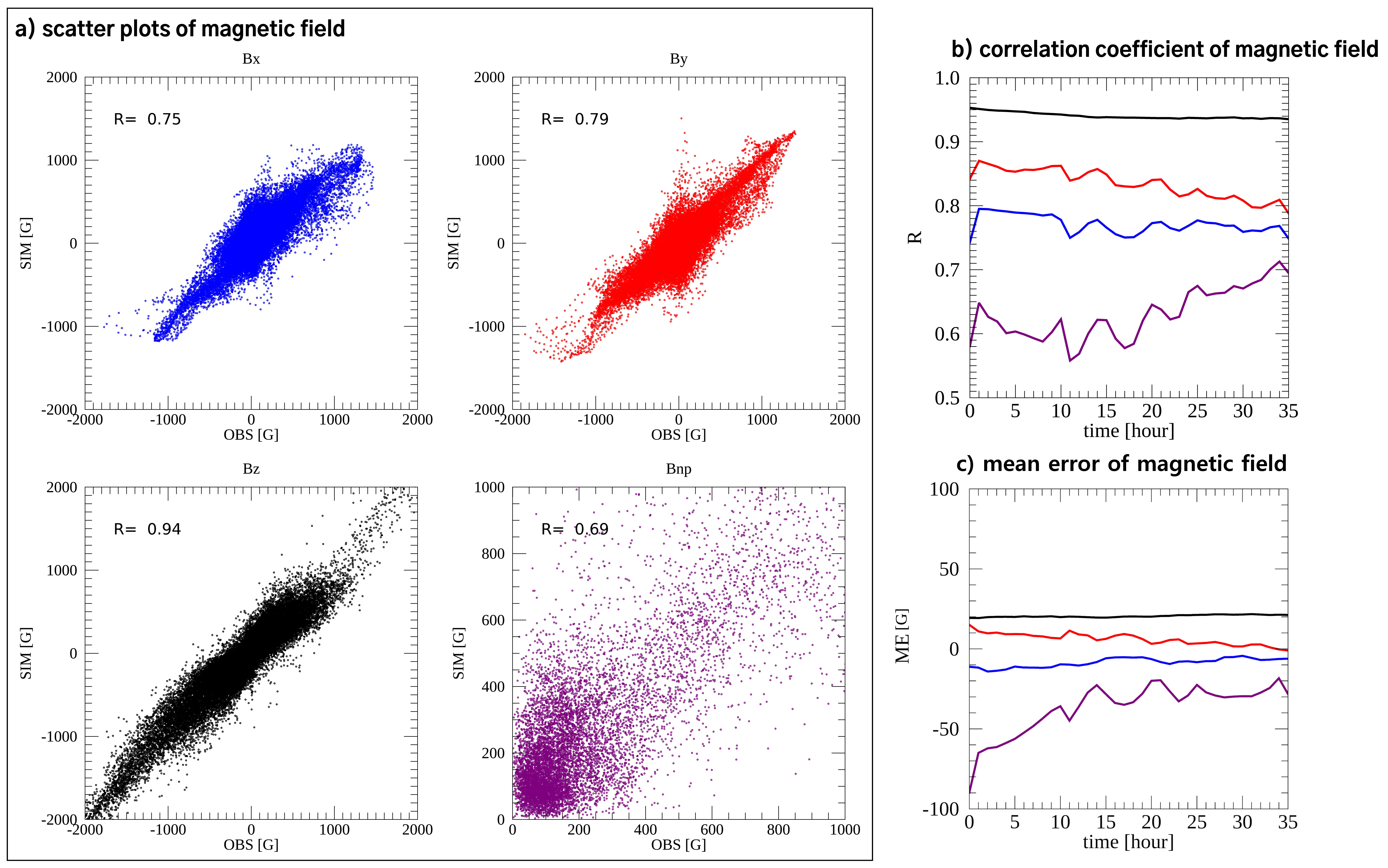}
\caption{Correlations between the observational magnetic fields and the simulated ones. (a) The scatter plots between the observational photospheric magnetic field ($B_{x}$, $B_{y}$ and $B_{z}$: within the FOV of Figure \ref{fig:bznp} and $B_{\mathrm{np}}$: within the red-boxed region in Figure \ref{fig:bznp}) and the simulated one at t=35 hr. Note that the simulated magnetic field was multiplied by a factor of 50 to compare it against the original magnitude. (b) The temporal evolution of correlation coefficients $R$ of $B_{x}$, $B_{y}$, $B_{z}$ and $B_{\mathrm{np}}$. (c) The temporal evolution of mean errors of $B_{x}$, $B_{y}$, $B_{z}$ and $B_{\mathrm{np}}$. The line color indicates the evolution of each component ($B_{x}$: blue, $B_{y}$: red, $B_{z}$: black and $B_{\mathrm{np}}$: purple).}
\label{fig:correlation}
\end{figure}

To confirm the reproducibility of our simulation, we investigated the scatter plots distributions for all the components of photospheric magnetic field within the FOV of Figure \ref{fig:bznp} and the non-potential field within the red-boxed region between the observation and the simulation results at t=35 hr, the terminated time of simulation. We also investigated the temporal evolution of their correlation coefficients $R$ during the simulation period. Figure \ref{fig:correlation} (a) presents the scatter plots of each component of magnetic field. Figure \ref{fig:correlation} (b) shows the temporal evolution of $R$. Shown in Figure \ref{fig:correlation} (a), the reproducibility of $B_{\mathrm{np}}$ was weaker than those for the magnetic field components $B_{x}$, $B_{y}$ and $B_{z}$. However, Figure \ref{fig:correlation} (b) shows that $R$ of $B_{\mathrm{np}}$ entirely increased from t=12 hr from $R=0.55$ to $0.70$. In addition, we investigated the mean error ($ME$) of magnetic field, and Figure \ref{fig:correlation} (c) displays the temporal evolution of mean errors. At the beginning of simulation, the mean error of $B_{\mathrm{np}}$ reached to -100 G, however, the mean error became small gradually finally reaching less than -50 G at t=35 hr. It indicates that the simulation can reasonably reproduce the non-potentiallity of this region. The reproducibility of the simulation could be considered acceptable, as we used the mask function during the inversion of the electric field with respect to the magnetic field. This approach indicates that magnetic fields weaker than 200 G (which corresponds to 4 G in the simulated field scale) were not consistently reproduced in quantitative terms.

To understand the flare-related energy evolution quantitatively, we investigated the evolution of several parameters including $E\_{injection}$ (Equation \ref{eq:energy injection}) over the HiFER as follows:

\begin{equation}
    \Phi\_{un}=\int |B_{z}|dA,
    \label{eq:Bz unsigned flux}
\end{equation}

\begin{equation}
    E\_{m}=\int \frac{B^{2}}{8\pi}dV,
    \label{eq:magnetic energy}
\end{equation}

\begin{equation}
    E\_{pot}=\int \frac{B\_{pot}^{2}}{8\pi}dV,
    \label{eq:potential energy}
\end{equation}

\begin{equation}
    E\_{free}=\int \frac{B^{2}}{8\pi}dV-\int \frac{B\_{pot}^{2}}{8\pi}dV,
    \label{eq:free energy}
\end{equation}

\begin{equation}
    E\_{k}=\int \frac{\rho v^{2}}{2} dV,
    \label{eq:kinetic energy}
\end{equation}

\begin{equation}
    \Phi_{\mathrm{twist}}=\int \tau dA=\int B_{z} T_{w}dA \quad \mathrm{for} \quad B_{z} > 0 \quad \mathrm{and} \quad T_{w} \geq T_{w \mathrm{th}}
    \label{eq:magnetic twist flux}
\end{equation}

\noindent
where $T_{w}$ is the magnetic twist number \citep{2006JPhA...39.8321B}, the total amount of twist with respect to the axis field line, given by $T_{w}= (1/4\pi) \int (\vctr{B}\cdot\nabla\times\vctr{B})/B^{2} dl$ where $\int dl$ is a line integration along a magnetic field line. $\tau$ is the magnetic twist flux density \citep{2020Sci...369..587K} given by $\tau=B_{z} T_{w}$. $T_{w\mathrm{th}}$ is a threshold of $T_{w}$ to define an MFR in our model. $\Phi\_{un}$ denotes the vertical unsigned magnetic flux. $E\_{m}$, $E\_{pot}$, $E\_{free}$ and $E\_{k}$ represent the magnetic energy, the potential energy, the free magnetic energy and the kinetic energy, respectively. $\Phi_{\mathrm{twist}}$ denotes the magnetic twist flux. We conducted both the surface integration ($\int dA$) in the red-boxed region encompassing the HiFER on the photosphere ($z=0$) , as in Section \ref{sec:model}, and volume integration ($\int dV$) over the red-boxed region ($z \geq 0$). We investigated $\Phi\_{un}$ for the observation and the input in the same position and the size of the red-boxed region. $\Phi_{\mathrm{twist}}$ was calculated under the two conditions: A positive magnetic flux should exist because the HiFER was mainly located over P1, and the magnetic twist numbers of the field lines from the positive patch are greater than $T_{w \mathrm{th}}=$ 0.60, 0.80 and 1.00. Thus, $\Phi_{\mathrm{twist}}$ includes information on both of (1) how much the magnetic field is strong, and (2) how much the field lines are twisted. The purpose for the set of $T_{w \mathrm{th}}$ is to find which $T_{w}$ is most appropriate to define the flare-related (or eruption-related) MFR and display their formation in our model.

\begin{figure}

\centerline{
\includegraphics[width=0.5\textwidth,clip,trim=50mm 25mm 0mm 20mm, angle=0]{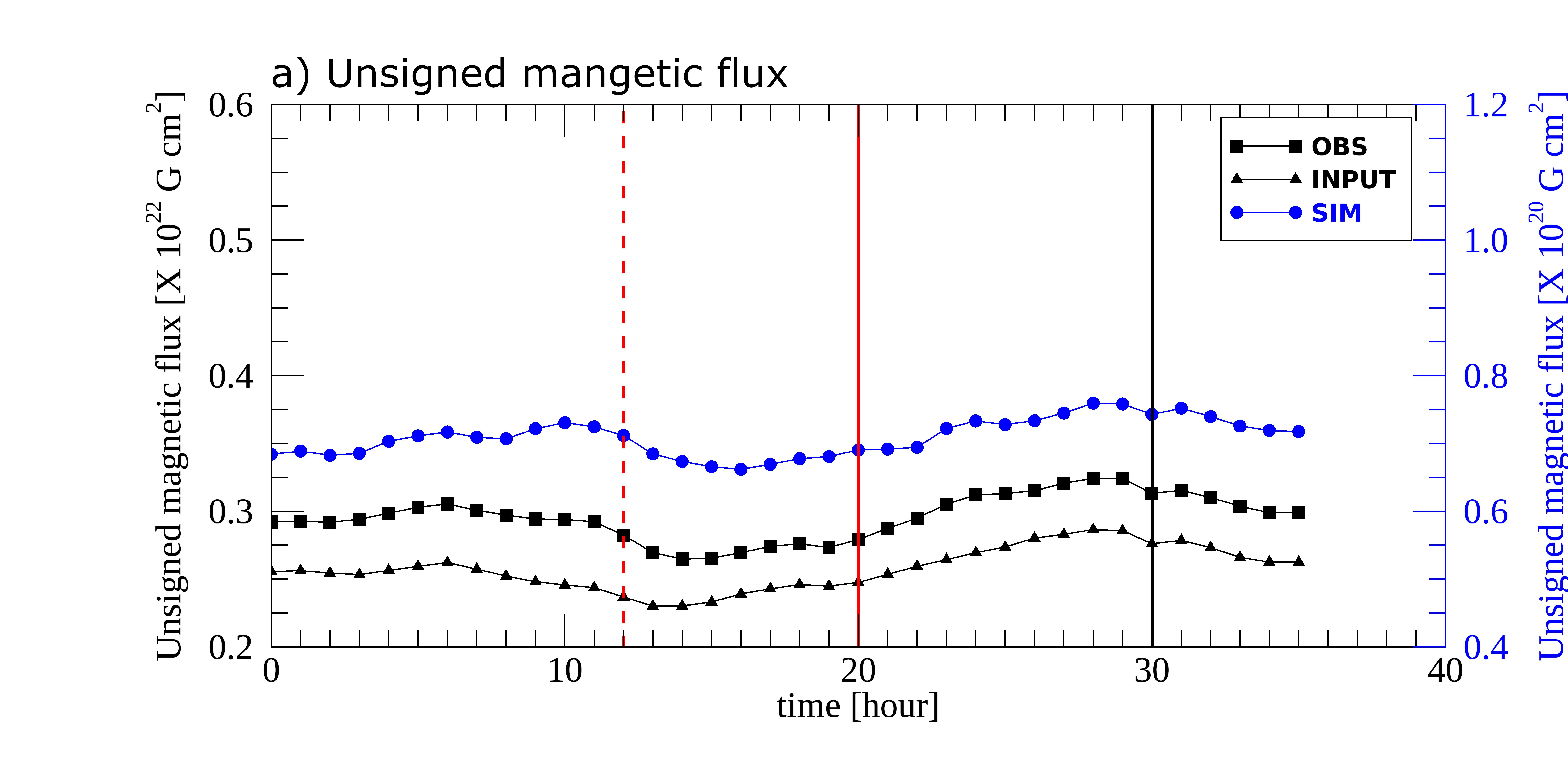}
\includegraphics[width=0.5\textwidth,clip,trim=50mm 25mm 0mm 20mm, angle=0]{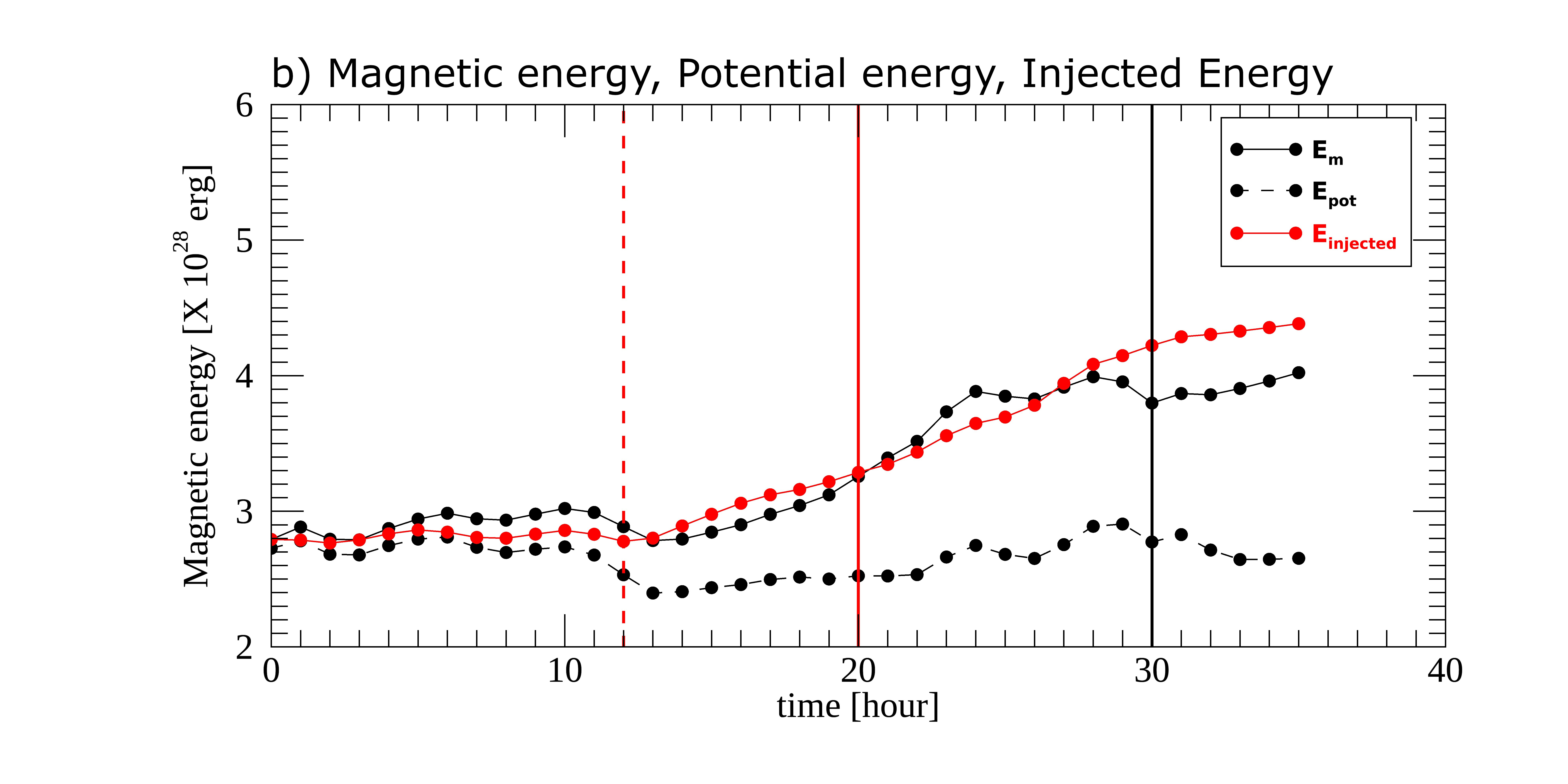}}
\centerline{
\includegraphics[width=0.5\textwidth,clip,trim=50mm 25mm 0mm 20mm, angle=0]{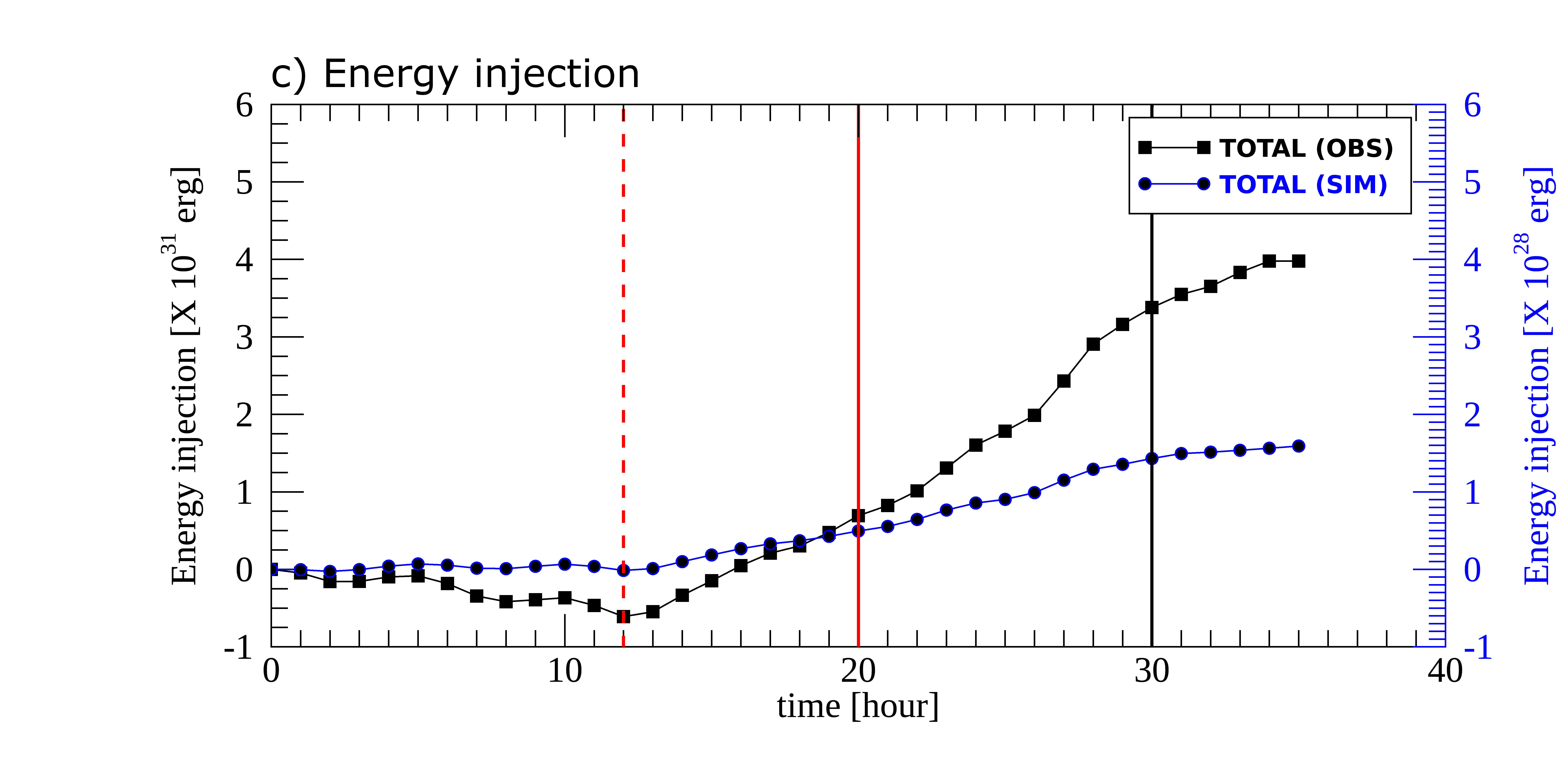}
\includegraphics[width=0.5\textwidth,clip,trim=50mm 25mm 0mm 20mm, angle=0]{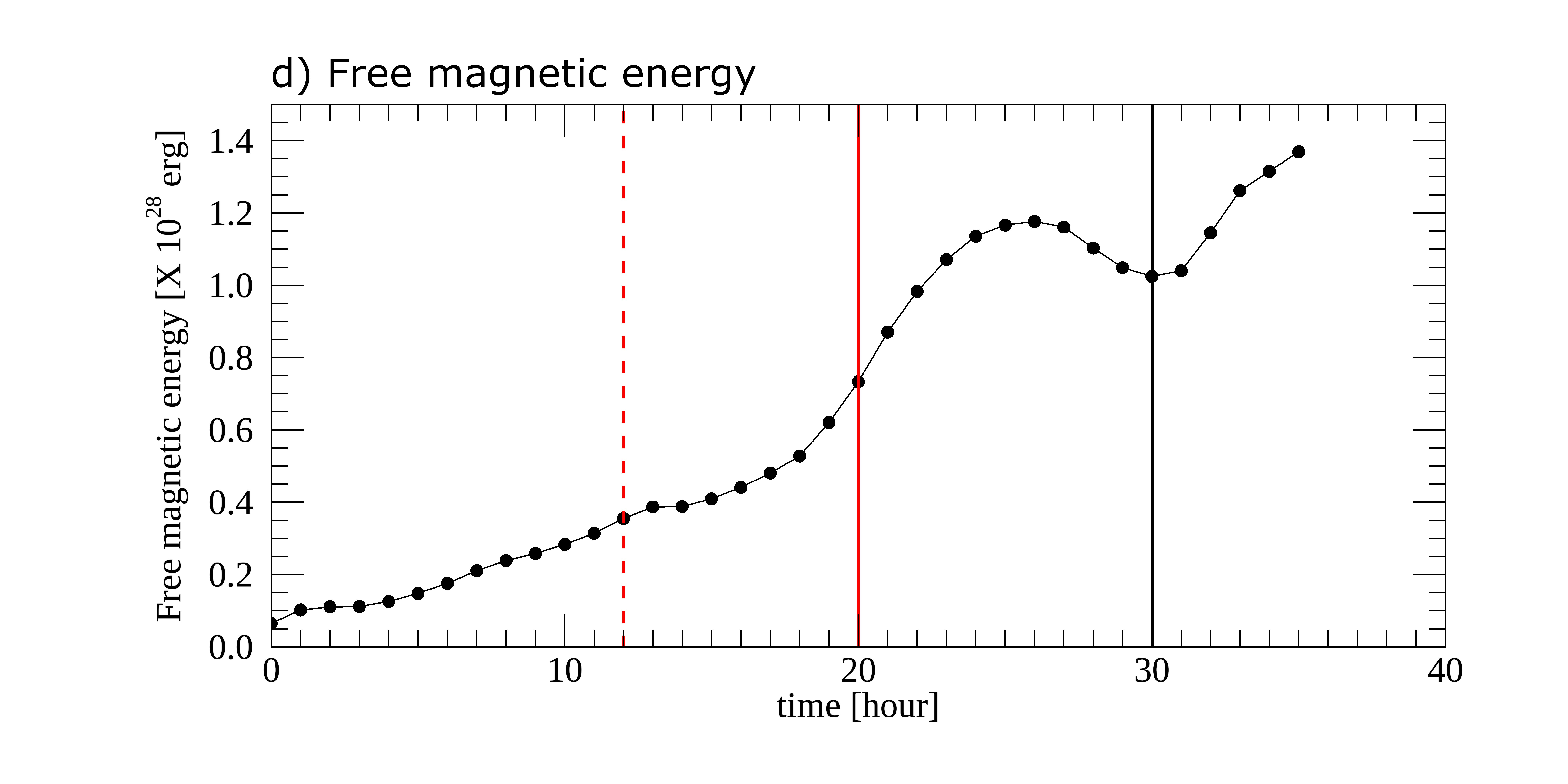}}
\centerline{
\includegraphics[width=0.5\textwidth,clip,trim=50mm 25mm 0mm 20mm, angle=0]{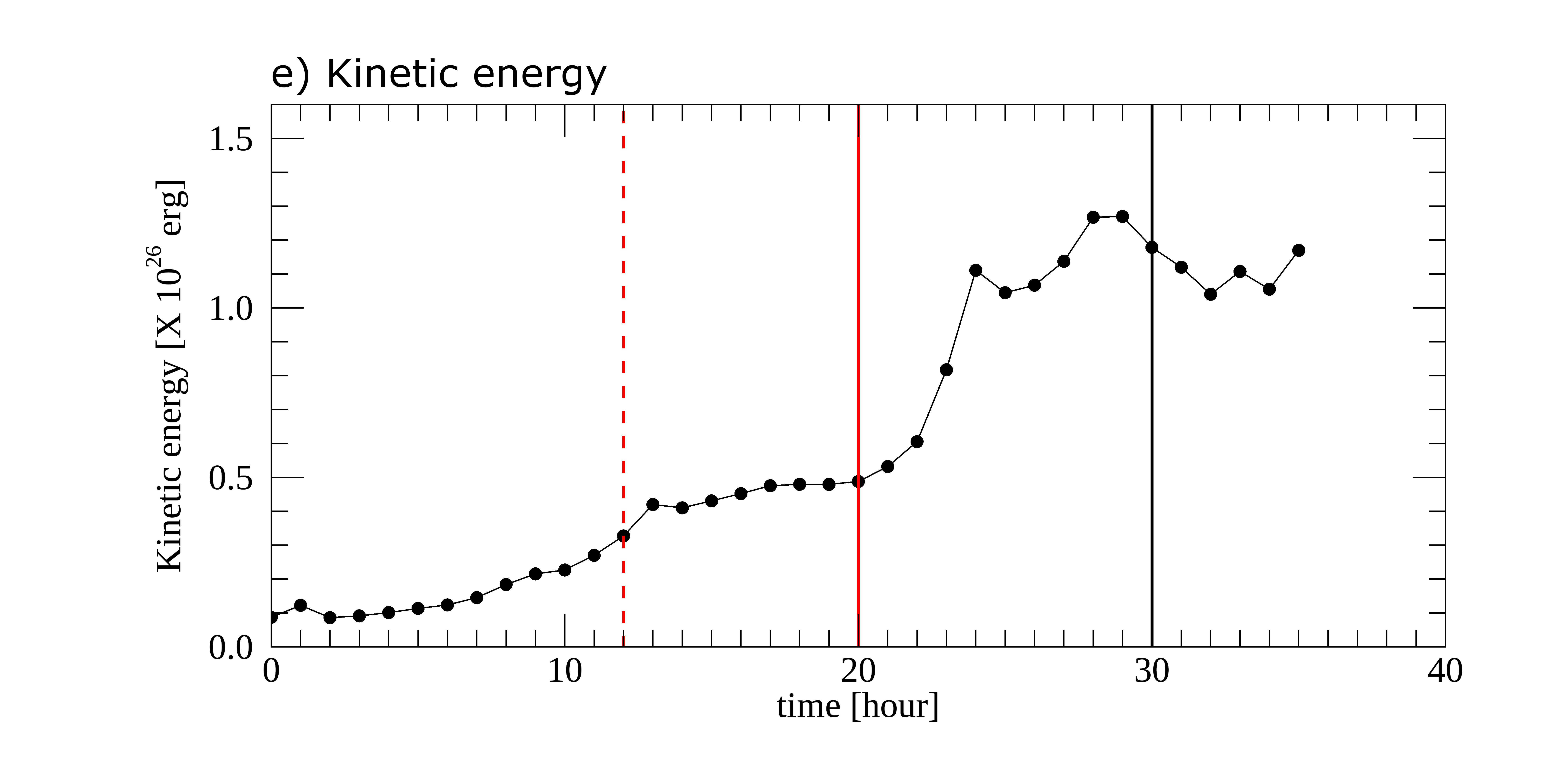}
\includegraphics[width=0.5\textwidth,clip,trim=50mm 25mm 0mm 20mm, angle=0]{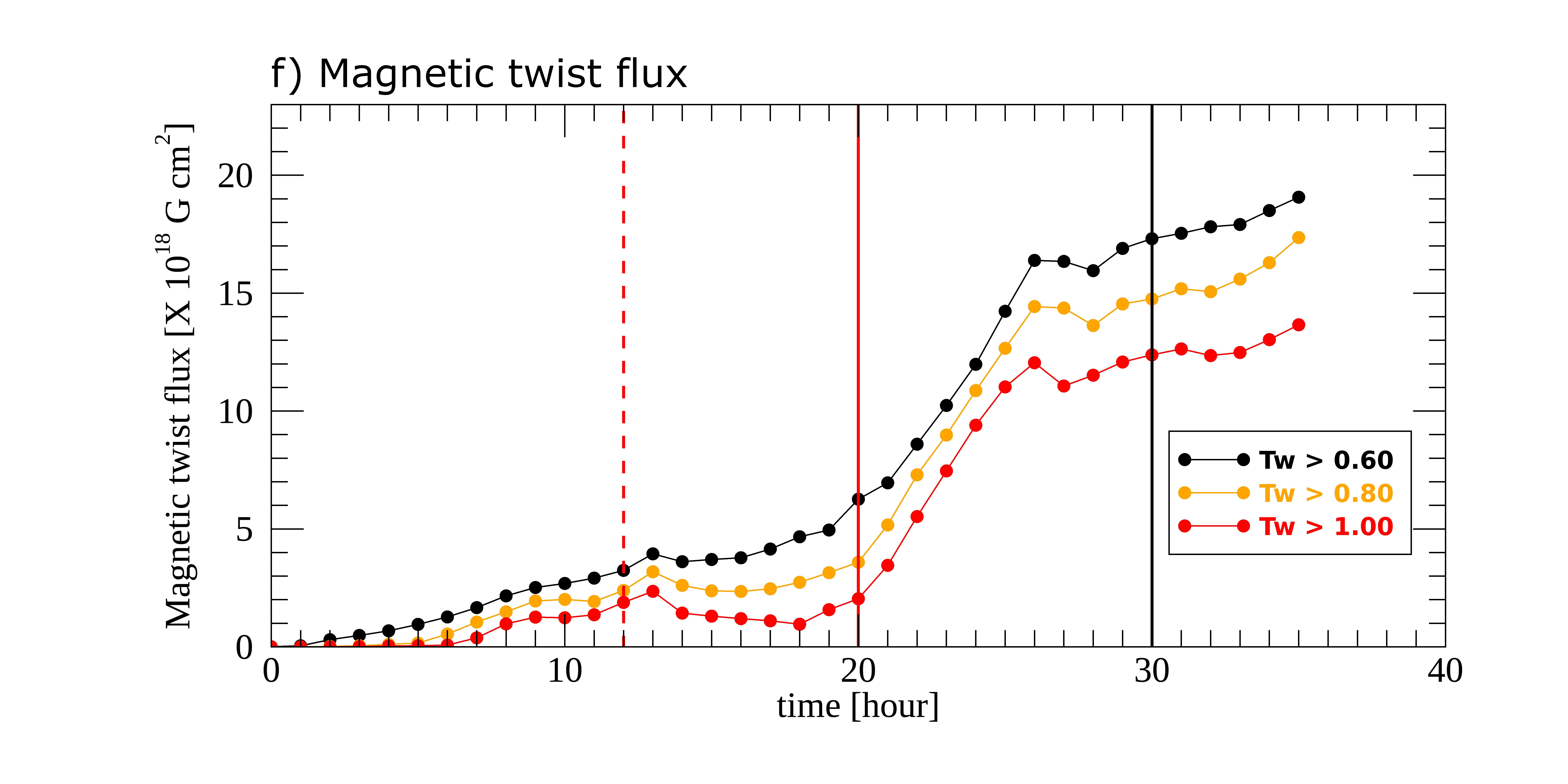}}
                
\caption{The temporal evolution of physical quantities over the HiFER, indicated by the red box in Figure \ref{fig:bznp}. (a) $\Phi\_{un}$ in the observation, input and simulated result, (b) $E\_{m}$ (solid), $E\_{pot}$ (dash) and $E\_{injection}$ (red), (c) $E\_{injection}$ in the observation and the simulation result, (d) $E\_{free}$, (e) $E\_{k}$ and (f) $\Phi_{\mathrm{twist}}$ with the different thresholds of $T_{w}$, $T_{w\mathrm{th}}=$ 0.60 (black), 0.80 (orange) and 1.00 (red). The vertical red solid (dashed) lines indicate the onset time of the reproduced full (failed) eruption, both exhibiting rapid increases of $E\_{k}$ in panel (e). The black solid lines in all the panels correspond to the SXR peak time of the M5.3 flare.}
\label{fig:temporal_quantities}    
\end{figure}

Figure \ref{fig:temporal_quantities} shows the temporal evolution of these parameters. The time notation follows the observational time scale, that is, t=0 corresponds to the observational time of September 4, 19:48 UT, 2011 (the initial condition of the simulation), and the time interval is 1 hour. Figure \ref{fig:temporal_quantities} (a) shows $\Phi\_{un}$ of the observation, the input and the simulation result. One can see that the overall pattern of the simulated magnetic field resembles the observed data, although there were some differences due to the influence of the low pass filter. After applying the low-pass filter, the overall magnitude of the unsigned magnetic flux decreases by 16\% compared to the observed data. The discrepancies, with a 40 times difference in the entire magnitude between the simulation and both the observed and input data, arise from reducing the magnetic field strength by a factor of 50 in the simulation. Figure \ref{fig:temporal_quantities} (b) shows the temporal evolution of $E\_{m}$, $E\_{pot}$ and the injected energy from the initial (t=0) $E\_{m}$. The changes in $E\_{pot}$ appears to be correlated with the simulated $\Phi\_{un}$ shown in Figure \ref{fig:temporal_quantities} (a), and they exhibit a good match. The temporal evolution of $E\_{m}$ was also roughly consistent with the injected energy.

\begin{figure}[htbp!] 
\centering    
\includegraphics[width=0.40\textwidth]{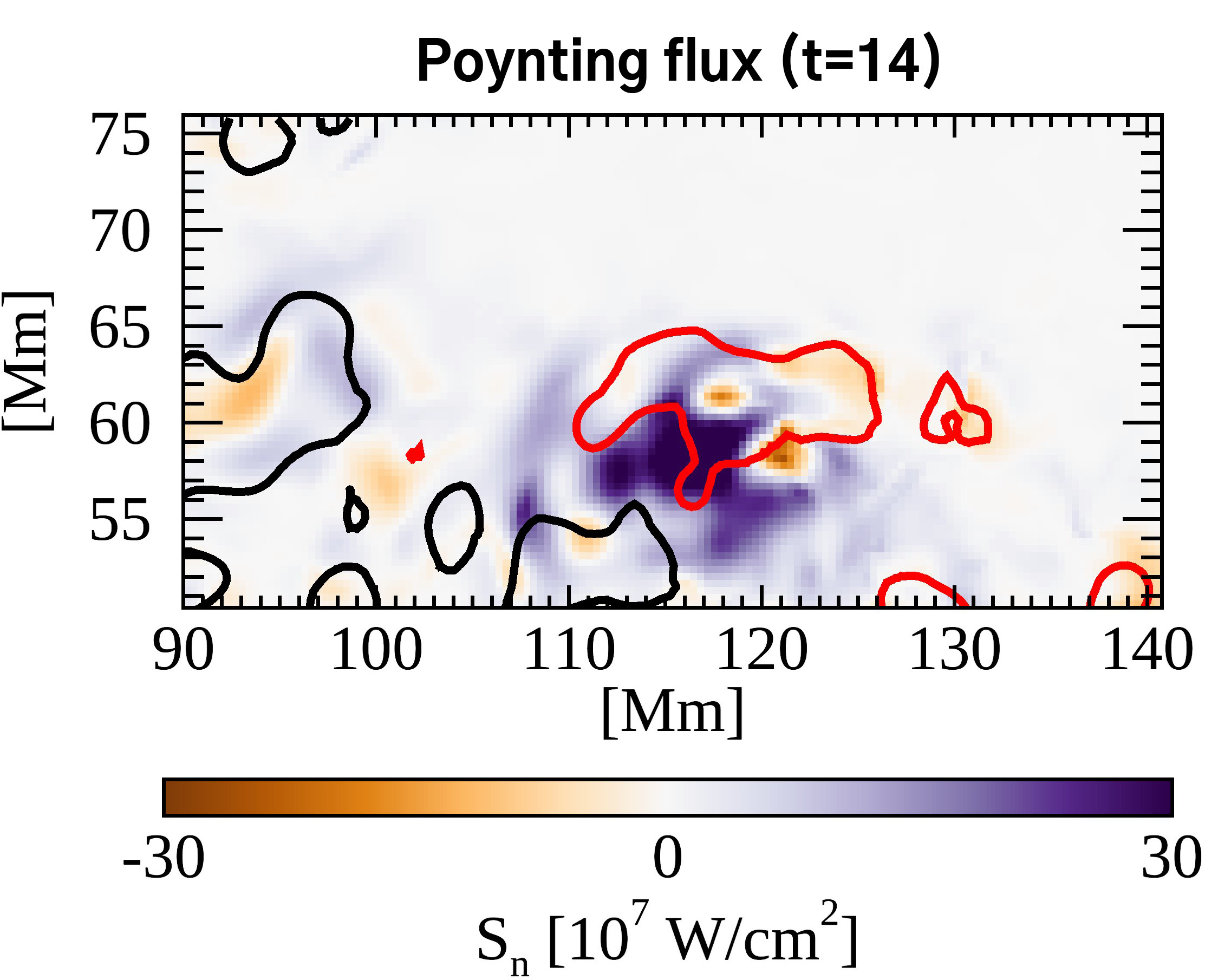}
\caption{The distribution of Poynting flux ($S_{n}$) on the photosphere in the FOV of Figure \ref{fig:bznp_mag} at t=14 hr. The solid red (black) line contours indicate $B_{z}=10$ G ($-10$ G), respectively (refer to the white and black line contours in Figure \ref{fig:bznp_mag} (b).)}
\label{fig:injection}
\end{figure}

Figure \ref{fig:temporal_quantities} (c) shows the comparison of $E\_{injection}$ (see Equation \ref{eq:energy injection}) in the observation and the simulation results. We used the photospheric velocity field obtained from our data-driven algorithm, same as in Section \ref{sec:model}. One can see that the trend of simulation results was similar to that of observation with a slight difference in the quantity of energy injection. For instance, in the period of t=12--30 hr, the energy injection in the observation was approximately $4.0 \times 10^{31}$ erg, whereas the energy injection in the simulation amounted to about $1.5 \times 10^{28}$ erg. Taking into account that the strength of the simulated magnetic field was reduced by a factor of 50 compared to the observational field (which means the energy was reduced by 2500 times), the energy injection on the observational scale amounted to about $3.75 \times 10^{31}$ erg. Therefore, this deviation seems acceptable. Figure \ref{fig:injection} presents the distribution of photospheric Poynting flux $S_{n}$ within the HiFER-focused region at t=14 hr, representing one moment of the primary energy injection. We found that the Poynting flux concentrated in the area of $115 \; \mathrm{Mm} <x<125$ Mm and $55 \; \mathrm{Mm} <y<60$ Mm, close to the location of PIL. This distribution suggests that the primary energy injection occurred within a smaller area between the positive and negative poles in the HiFER.

Figure \ref{fig:temporal_quantities} (d) and (e) represent the temporal evolution of $E\_{free}$ and $E\_{k}$, respectively. In Figure \ref{fig:temporal_quantities} (d), $E\_{free}$ was gradually accumulated from the beginning of the simulation with one manifest decrease in the period of t=26--30 hr. In Figure \ref{fig:temporal_quantities} (e), in particular, two rapid increases of $E\_{k}$ appear at t=12 and 20 hr, and their subsequent variations exhibited a different trend. The evolution of $E\_{k}$ showed a small increase after t=12 hr with no more subsequent increase during t=15--19 hr.  After t=20 hr, on the other hands, a manifest increase was exhibited. Considering these different trends, we defined the first small increasing phase as a failed eruption (t=12--14 hr) and the second increasing phase as a full eruption (t=20--29 hr). Although the growth of $E\_{k}$ should be due to the conversion from the magnetic free energy $E\_{free}$, $E\_{free}$ did not start to decrease at the onset of either eruption. We explain this as being due to the different orders of magnitude of the two energy terms, with $E\_{k}$ being substantially smaller than $E\_{free}$.

Figure \ref{fig:temporal_quantities} (f) represents the temporal evolution of $\Phi_{\mathrm{twist}}$ with different $T_{w \mathrm{th}}$. One can see that the overall trends of $E\_{free}$ and $\Phi_{\mathrm{twist}}$ for all $T_{w \mathrm{th}}$ values were consistent with each other before t=27 hr. The comparison of the temporal evolution of $E\_{k}$ and the different growth rates of $\Phi_{\mathrm{twist}}$ suggests that the definition and identification of an MFR is necessary to identify twisted field lines associated with the eruptions. To explain the quantitative evolution in both the eruption phases, we define an MFR as magnetic field lines whose one-side footpoints were located near P1 included in the HiFER and have a total twist given by $T_{w}>0.80$. Comparing to the observational time, however, the production of the full eruption was a few hours earlier than the observational time of the M5.3 flare peak.

\begin{figure}[htbp!] 
\centering    
\includegraphics[width=0.80\textwidth]{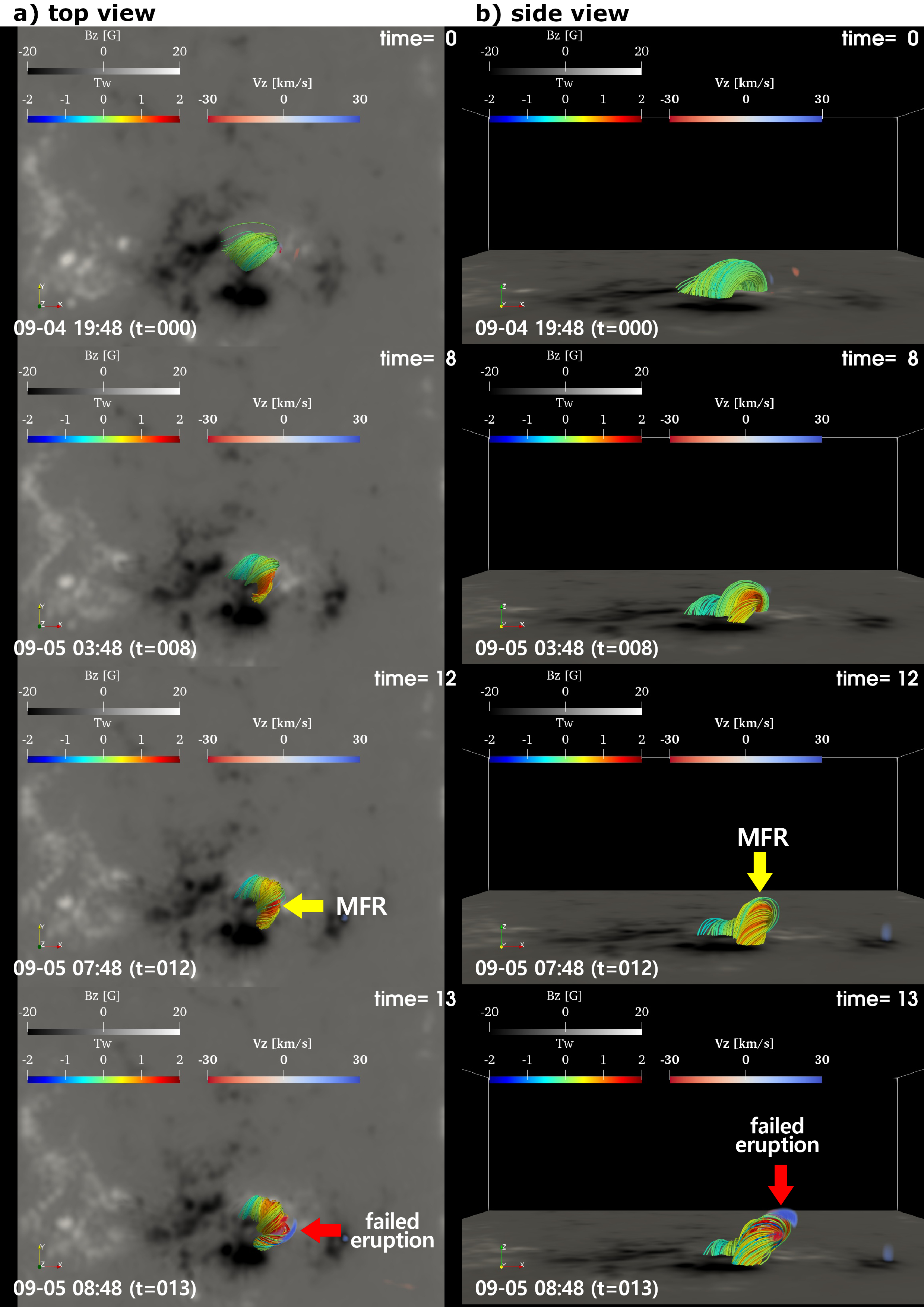}
\caption{Formation of the MFR (t=0--8 hr) and the failed eruption (t=12--13 hr) reproduced in the simulation. Panels (a) and (b) show the temporal evolution of the three-dimensional magnetic field from the vertical view and the side view (along the $y$-axis), respectively. The bottom boundary displays the simulated vertical magnetic field ($B_{z}$) on the photosphere with a limited FOV. The color of the coronal magnetic field lines indicates $T_{w}$. The volume rendering corresponds to the distribution of the vertical velocity ($v_{z}$). The yellow arrows indicate the MFR. The failed eruption is indicated by the red arrows and corresponds to the region where high velocity exists.}
\label{fig:3d_failed}
\end{figure}

\begin{figure}[htbp!] 
\centering    
\includegraphics[width=0.80\textwidth]{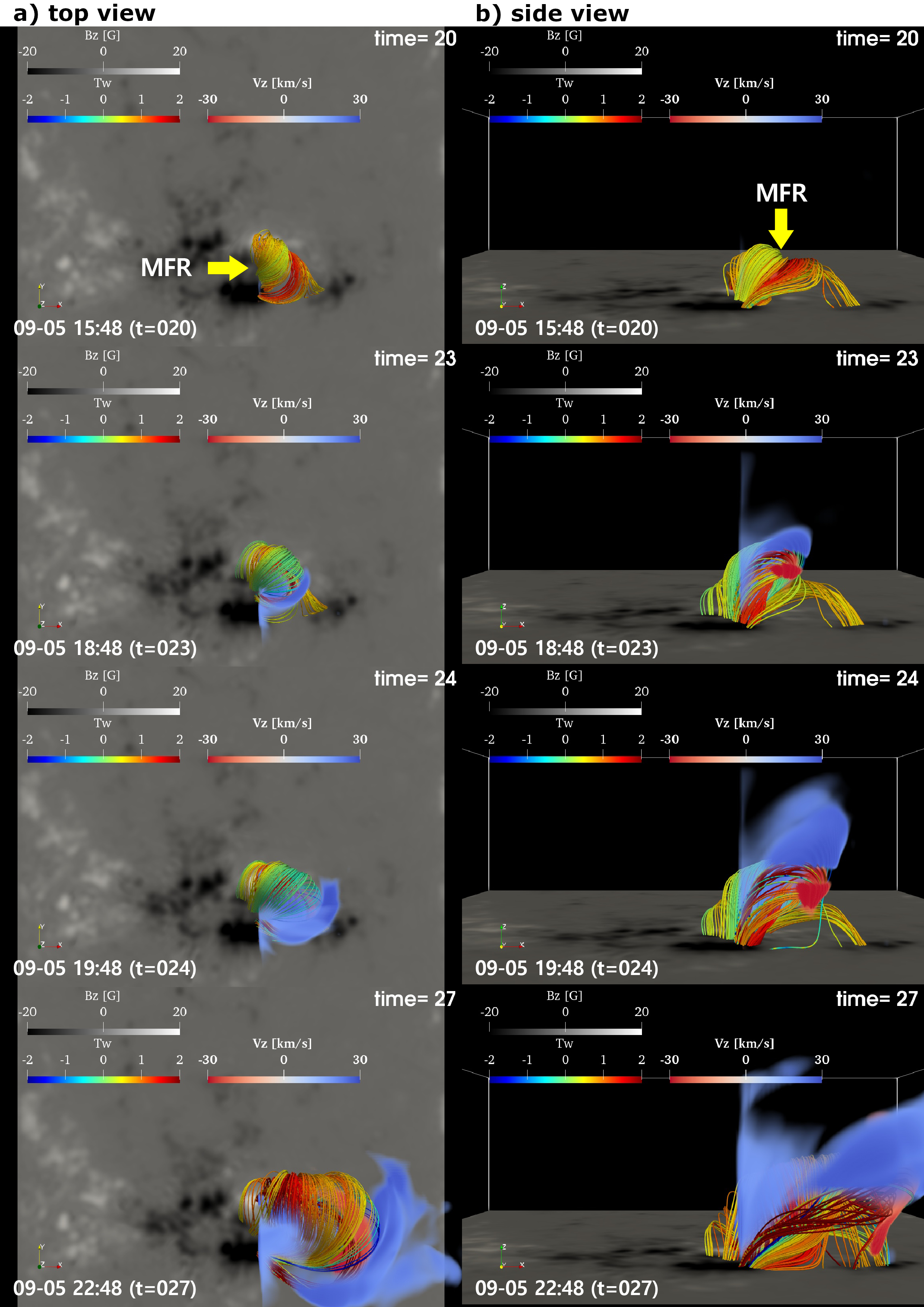}
\caption{The full eruption corresponding to the M5.3 flare (t=20--29 hr). Panels (a) and (b) show the erupting motion of the MFR from the same viewpoint as Figure \ref{fig:3d_failed}. The yellow arrows indicate the MFR at the onset of the full eruption. The full eruption is inclined from the $z$-direction. Note that the images only display the velocity rendering at the region of $108$ Mm $<y<176.4$ Mm for the better illustration of the full eruption.}
\label{fig:3d_full}
\end{figure}

Our modeling produced a three-dimensional evolution of simulated magnetic field, exhibiting the formation of an MFR, a failed eruption and a full eruption which corresponds to the observed M5.3 flare. Figure \ref{fig:3d_failed} shows both the MFR formation and the following failed eruption. Figure \ref{fig:3d_full} displays the full eruption with same FOV as Figure \ref{fig:3d_failed}. 

Regarding the failed eruption, the magnetic field was almost potential as in the initial condition (t=0) but some twisted field lines were formed by the convergence of P1 into N2 from t=8 hr on, leading to the formation of the MFR. At t=13 hr, the three-dimensional evolution showed a small high velocity distribution near the apex of the MFR, which can be interpreted as the failed eruption. This result was also consistent with the evolution of $E_{\mathrm{k}}$ shown in Figure \ref{fig:temporal_quantities} (e). After the failed eruption, the MFR  stabilized from t=15--19 hr then started to erupt again at t=20 hr corresponding to the onset time of the full eruption (Figure \ref{fig:3d_full}). From both figures, it is evident that the MFR erupted at an inclination relative to the vertical direction. We will analyze the mechanism of this inclined eruption in Section \ref{subsec:torus_instability}.

However, this modeling was not able to reproduce the complete ideal relationship between $E_{\mathrm{free}}$ and $E_{\mathrm{k}}$ during the eruptions because the energy equation was not solved in the simulation, resulting in a loss of Joule heating. In terms of the monotonic increase of $E_{\mathrm{free}}$ during the full eruption, one possible reason is that $E_{\mathrm{injection}}$ was much greater than the decrease of $E_{\mathrm{free}}$ during the eruption. After the full eruption, the slight decrease of $E_{\mathrm{free}}$ may be attributed to the deviation of the field lines in the erupting MFR from the analyzed region. This could be because the integration area was too confined in comparison to the broader simulation domain. Moreover, the evolution of $E_{\mathrm{free}}$ and $E_{\mathrm{k}}$ can be much slower than the reality because we assumed the slower Alfvén speed in the simulation.

\begin{figure}[htbp!] 
\centering    
\includegraphics[width=0.9\textwidth]{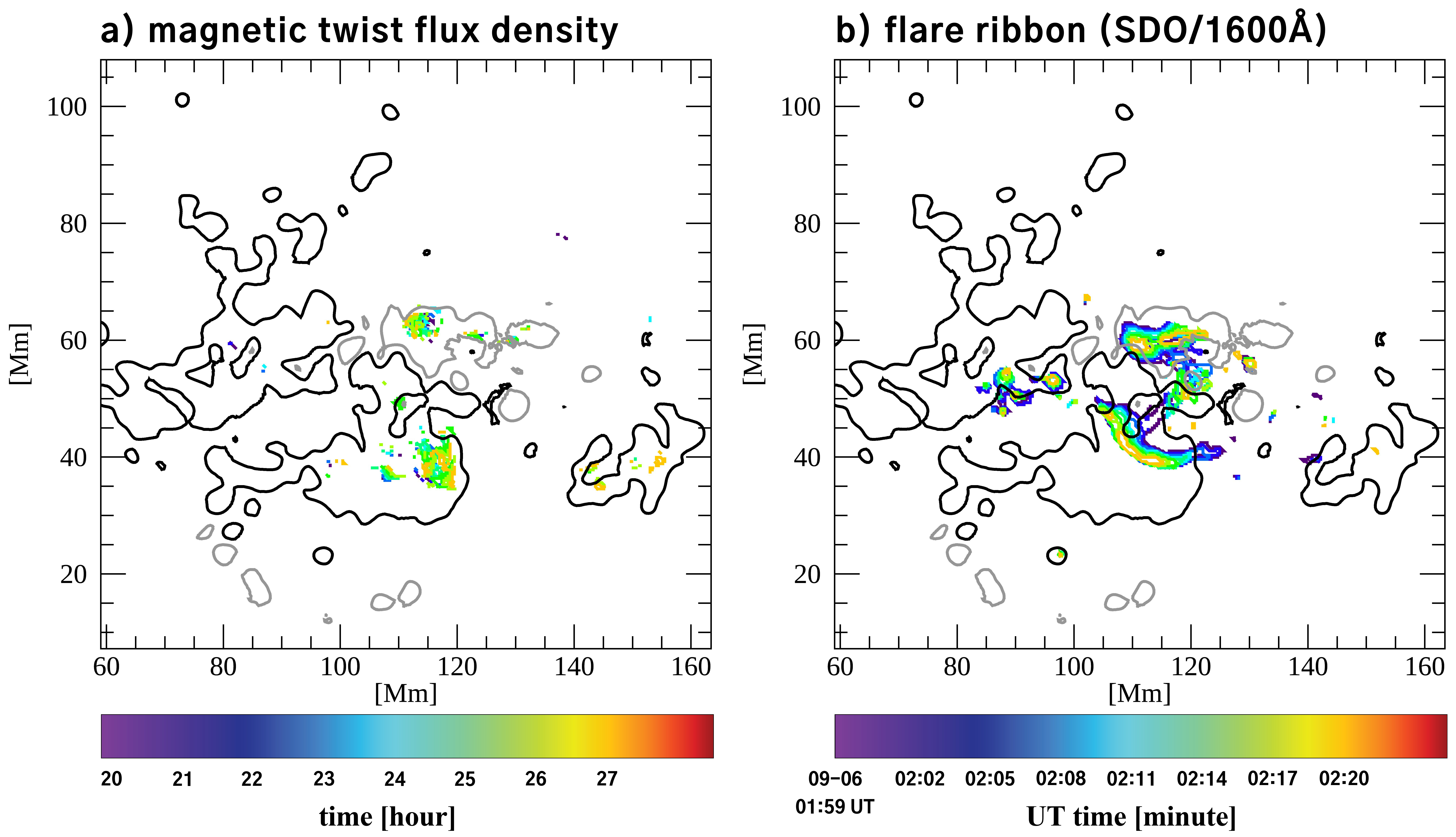}
\caption{Comparisons of the simulation results from t=20--27 hr (panel a) the SDO/AIA 1600 {\AA} observations from September 6 01:59 UT, 2011 (the flare peak time) to September 6 02:20 UT, 2011 (panel b). In panel (a), the time t=20 hr corresponds to the onset of full eruption in the simulation. The flare ribbons are defined using the same method outlined in Section \ref{sec:event}. The gray and black line contours in both panels show the photospheric vertical magnetic field ($B_{z}$ at +/- 20 G, which will be equivalent to +/- 1000 G in the original scale) from the simulation at t=20 hr (refer to the white and black line contours in Figure \ref{fig:bznp_mag} (b), or the red and black line contours in Figure \ref{fig:injection}). The colored line contours in panel (a) show represent the the magnetic twist flux density ($\tau$) increase of 15 G. The colored line contours in panel (b) shows the temporal changes of the flare ribbon distributions. The time interval of flare ribbon projections is 3 minutes in the observational time scale, while it is 1 hour in the simulation time scale.}
\label{fig:tw1600}
\end{figure}

In order to confirm a level of agreement between the simulation and observations, we compared the simulation to the SDO/AIA 1600 {\AA} image sequences of the M5.3 flare, guided by several studies that have suggested basic physical models of solar eruptions. The CSHKP model (\citealt{1964NASSP..50..451C}; \citealt{1966Natur.211..695S}; \citealt{1974SoPh...34..323H}; \citealt{1976SoPh...50...85K}), has suggested that solar eruptions can be driven by magnetic reconnections
of field lines at or near the solar surface, which
then generate energetic particles at the reconnection sites which can then precipitate into the chromosphere and are observed as flare ribbons. For example, considering a scenario where an MFR is positioned over a PIL, when the MFR starts to erupt field lines within
the erupting MFR can reconnect with its neighboring field lines along the eruption path. If these neighboring field lines possess positive helicity, they can contribute to a helicity increase in the MFR through the reconnections. Observationally, the eruption-related reconnection causes the precipitation of high-energy particle into the footpoints of the MFR, which manifests as flare ribbons resulting from chromospheric evaporation.

However, our simulation cannot reproduce the particle
acceleration and the chromospheric evaporation. Therefore, we analyze the footpoints of reconnected magnetic field lines in the simulation by way of the change in the magnetic twist flux density ($\tau$), and compare those areas to the flare ribbons observed in the SDO/AIA 1600 {\AA} image sequences, assuming that the magnetic twist must change as the result of magnetic reconnection. The test proceeded as follows: we calculated the difference of magnetic twist flux density ($\Delta \tau$) between each hour from t=20 hr (the onset time of full eruption) and its respective value 1 hour prior. Then, we compared the areas showing notable $\tau$ increases as a function of time, with corresponding SDO/AIA 1600 {\AA} image time series for approximately 20 minutes starting from September 6, 01:59 UT, 2011 (the flare peak time). Figure \ref{fig:tw1600} (a) shows the temporal evolution of $\tau$ differences ($\Delta \tau$) in the simulation, and the rainbow contours indicate $\Delta \tau=15$ G at different times. Figure \ref{fig:tw1600} (b) displays the temporal evolution of observed flare ribbon. We found regions in the model results that showed continual $\tau$ increases, implying that flare-related reconnection was occurring during the full eruption simulation. Comparing the observed flare ribbon and the $\Delta \tau$ evolution, they appeared on positive and negative poles adjacent to the same PIL, although the distribution of $\Delta \tau$ was more compact than the flare ribbon. It suggests that the simulated full eruption corresponds morphologically to the M5.3 flare. However, the change in $\Delta \tau$ related to the simulated eruption and flare-related reconnections occurred slightly earlier than in the observed flare. The difference is likely because the microscale dynamics and dissipation, which are difficult to reproduce accurately in simulations, could play a role in triggering the onset of flares.

\section{discussions} \label{sec:discussions}

\subsection{Failed {\it vs.} full eruptions} \label{subsec:failed_eruption}

\begin{figure}[htbp!] 
\centering    
\includegraphics[width=1.00\textwidth]{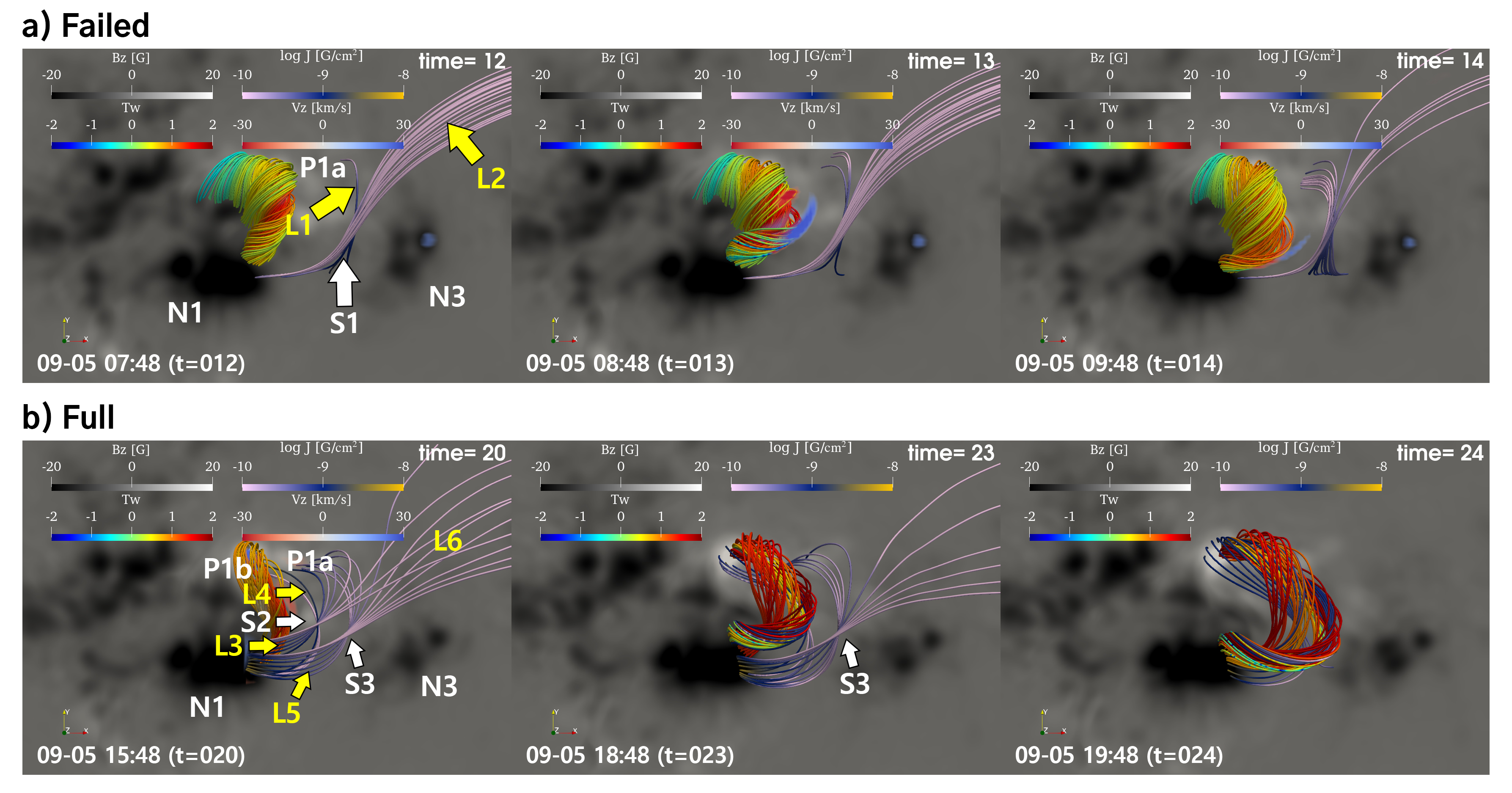}
\caption{Three-dimensional magnetic topology of the failed eruption (a) and the full eruption (b), showing the zoomed-in views of Figure \ref{fig:3d_failed} (a) and Figure \ref{fig:3d_full} (a), respectively. The white arrows (S1, S2 and S3) are points in the QSLs with lower field strength. The yellow arrows indicate groups of the magnetic field lines related to S1, S2 and S3. In panel (a), the field lines of L1 and L2 are related to S1: the L1 field lines connecting from P1a to N3, and the L2 field lines from the side-boundary to N1. In panel (b), the field lines labelled L3 and L4 are related to S2, with L3 field lines connecting P1a to N1, L4 connecting from P1b to N1. The L5 and L6 field lines are related to S3, with L5 field lines connecting P1a to N1, and L6 field lines from the lateral boundary to N1. The color of the QSL-related field lines indicate the total electric currents flowing along them on a logarithmic scale.}
\label{fig:qsl}
\end{figure}

\begin{figure}[htbp!] 
\centering    
\includegraphics[width=1.00\textwidth]{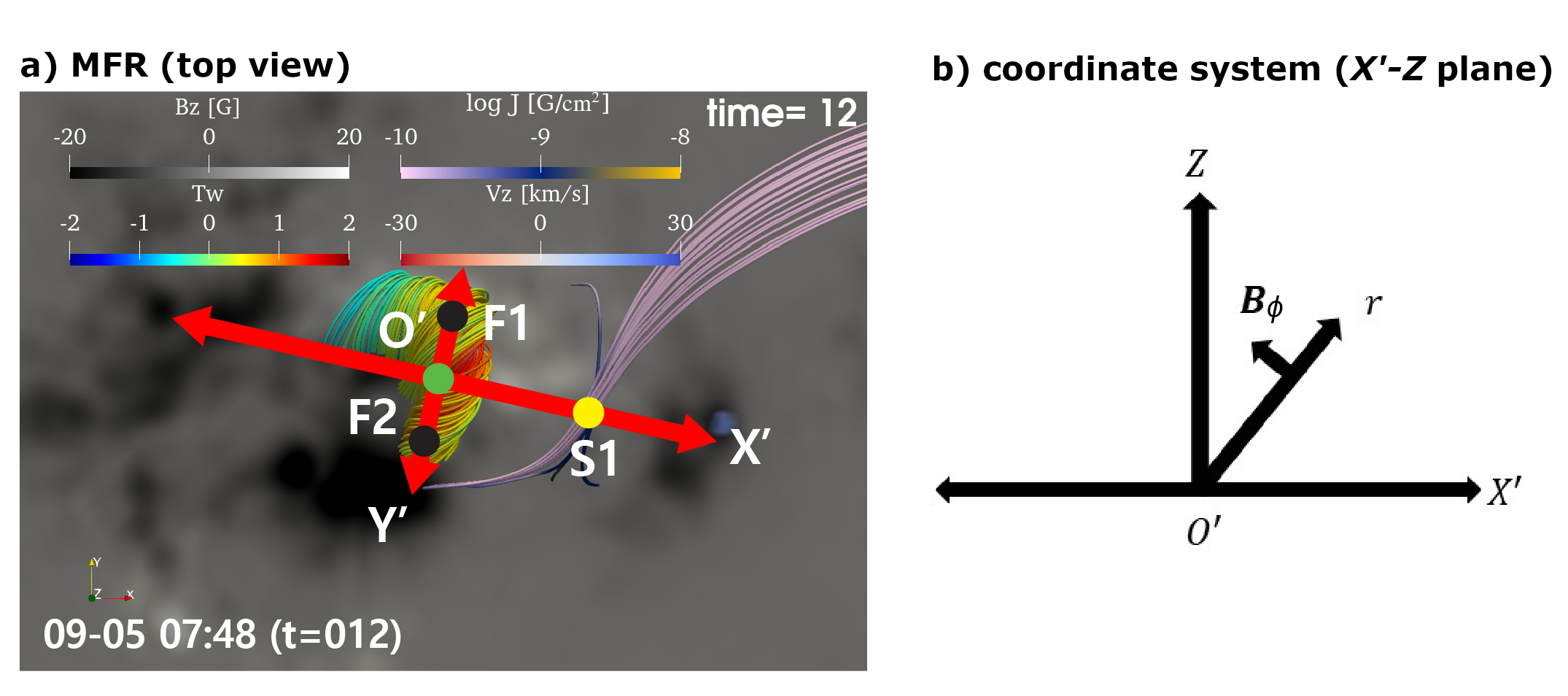}
\caption{Cylindrical coordinate system along the direction of the movement of the MFR. Panel (a) shows the vertical view of the MFR and the neighboring magnetic field lines related to S1 at t=12 hr. The background image in panel (a) shows the photospheric magnetic field ($B_{z}$). F1 and F2 (the black circles) are two footpoints of one magnetic field line included in the MFR. $O'$ (the green circle) is the center of the MFR when it is assumed as an ideal torus plasma. The yellow circle indicates S1 with the related magnetic field lines. Note that S1 does not lie on the $X'$-axis (on the photosphere) but is located in the $X'-Z$ plane (above the photosphere). Panel (b) shows the schematic drawing of the side view of the cylindrical coordinate system ($r$,$\phi$). The $Z$-axis is perpendicular to the photosphere, and the $r$-axis is an arbitrary axial-radial direction from $O'$, which indicates any eruptive directions. $B_{\phi}$ is assumed as the confining magnetic field in this coordinate system. This coordinate system is defined and fixed by the magnetic topology at t=12 hr.}
\label{fig:cylindrical}
\end{figure}

To understand the differences between three-dimensional magnetic topology of two eruptions, we investigated the neighboring field lines near the MFR. Figure \ref{fig:qsl} (a) shows the magnetic topology of the failed eruption from the onset time (t=12 hr). From t=12 hr, we divided P1 into two distinct magnetic patches and newly defined them as P1a and P1b. This is because P1 started to be separated in an east-west direction after t=12 hr, and the separating motions changed the configuration of the neighboring field lines. One can see that a part of a quasi-separatrix layer (QSL), which differentiates regions of different magnetic field connectivity, was located in the west side of the MFR. To find the QSL related to the failed eruption, we located one point (``S1'') vertically above the $X'$-axis shown in Figure \ref{fig:cylindrical} (We will explain the definition of this axis in detail in Section \ref{subsec:torus_instability}). At S1, a very weak magnetic field existed numerically because the connectivity changed dramatically. S1 was formed by the opposite photospheric motions of P1a and N3, two footpoints of the field lines L1. The temporal evolution of the MFR showed that its erupting motion tried to proceed toward S1 where low magnetic pressure existed, but the eruption was terminated at t=14 hr. 

Figure \ref{fig:qsl} (b) displays the magnetic topology of the later full eruption. At its onset, the S1-related QSL moved further to the west, possibly due to the westward inclining motion of MFR as a consequence of the failed eruption. Even though this QSL disappeared, two new QSLs (related to two new points ``S2'' and ``S3'', also located above the $X'$-axis) were formed, also located in the western portion of the MFR. The formation of these new QSLs was caused by splitting motions in P1a, that is, P1b was formed, followed by the formation of S2 and S3. In particular, S2 was located in the east side of the MFR, which was closer than the position of S1. This magnetic topology could provide the MFR with an environment to erupt more easily than the failed case because very weak magnetic pressure may exist at S2. Considering this topology, as as result, the MFR merged these QSLs related to S2 and S3 sequentially during the full eruption.

We could not find any observational evidence of the failed eruption. One possibility is that the eruption was too small to be observed directly. Alternatively, of course, the failed eruption may exist only in the simulation. It is not fully understood from the diagnostics performed here why the MFR was unable to erupt in the first period of increased kinetic energy, and instead stabilized after the failed eruption. One possibility may be a boundary effect (or a line tying effect) from the bottom boundary defined as the photosphere in the simulation. In terms of this effect, \citet{2015ApJ...814..126Z} has found that field lines tied at the photosphere can interrupt local eruptions. This effect could be driven by the reproduced photospheric magnetic fields having strong magnitude, compared to the coronal region because the MFR's footpoints were anchored at the photosphere. However, we could not find any direct evidence of the boundary effect.

\subsection{Kink instability}
\label{subsec:kink_instabilty}

It is basically accepted that the mechanisms of solar eruptions are related to the three-dimensional magnetic topology. One approach to explain the initiation of these explosive events is an MHD instability within a specific topology. One mode is the kink instability, which is driven by an electric current flowing along an MFR (\citealt{1958PhFl....1..421K}; \citealt{1966RvPP....2..103S}; \citealt{2005ApJ...630L..97T}). If an MFR is bent to top-side by the electric current, the magnetic pressure increases at up-side, meanwhile, it decreases at the down-side. As a result, the instability can continue to grow up, driven by the magnetic pressure difference. An eruption then can be initiated by a rising MFR, possibly exhibiting a deformation, for example, the transformation of twist into writhe (\citealt{2005ApJ...630L..97T}). To grow the kink mode, the threshold of magnetic twist number at MFR's footpoints is suggested that $T_{w}$ is at least greater than 1.0 (\citealt{1958PhFl....1..421K}; \citealt{1966RvPP....2..103S}; \citealt{1981GApFD..17..297H}). In our results, even though the magnetic twist number of the reproduced MFR exceeded 1.0 sufficiently in the erupting phase (see Figure \ref{fig:3d_full}), however, we could not find the deformation of MFR, which result could suggest that the eruption was not initiated by the kink mode.

\subsection{Torus instability}
\label{subsec:torus_instability}

Another possible mode of the MHD instability is the torus mode. In decades, the torus instability (\citealt{1966RvPP....2..103S}; \citealt{1978mit..book.....B}; \citealt{2006PhRvL..96y5002K}) , which is one mode of an MHD instability, has been suggested. Assuming an MFR is an ideal torus plasma with an imaginary current below the photosphere, with an external magnetic field that tries to confine the torus (MFR), the torus instability can grow and upset the balance between the hoop (upward/outward) force driven by the electric current flowing in the torus and the Lorentz (constraining) force provided by the external field. \citet{1978mit..book.....B} suggested that the possibility for a torus instability to grow can be evaluated by means of evaluating the vertical gradient of external magnetic field, called the decay index $n$ (hereinafter referred to as the vertical decay index) as follows:

\begin{equation}
    n=-\frac{z}{B_{\mathrm{ex}}}\frac{dB_{\mathrm{ex}}}{dz}
    \label{eq:decay_index}
\end{equation}

\noindent
where $B_{\mathrm{ex}}$ is the confining external magnetic field. The threshold of the vertical decay index is accepted to be $n_{\rm crit}=1.5$ \citep{2006PhRvL..96y5002K}, however, several studies (\citealt{2010ApJ...718.1388D}; \citealt{2010ApJ...718..433O}; \citealt{2014ApJ...789...46K}; \citealt{2015ApJ...814..126Z}) have suggested that the threshold is not unique, and may change depending on numerical schemes and boundary conditions. However, the vertical decay index may not be appropriate for the case of an eruption that is inclined with respect to the local normal direction, as we find for the M5.3 flare. 

To include in the analysis an inclination angle of the eruption, we consider an axial-radial decay index $n_{r}$ in which the direction of the computed magnetic field fall-off is along an axial-radial direction oriented from the center of the torus.  
Some previous studies (\citealt{2019ApJ...870L..21G}; \citealt{2021NatCo..12.2734Z}; \citealt{2021ApJ...909...91K}) have introduced a similar parameter to evaluate how sharply the external field decays along the previously-determined path of an inclined eruption. Their analyses indicate that even an inclined eruption can be understood in the context of the torus instability. However, these prior works did not explain what determines the inclined angle of eruption {\it a priori}.

Here, we seek to establish the direction of eruption in a pre-eruptive phase.  We have calculated the axial-radial decay index in all radial directions prior to the eruption.
Assuming a cylindrical coordinate system whose origin is located at the center of the torus, we defined the axial-radial decay index $n_{r}$ as follows:

\begin{equation}
    n_{r}=-\frac{r}{B_{\phi}}\frac{dB_{\phi}}{dr}
    \label{eq:radial_decay_index}
\end{equation}

\noindent
where $B_{\phi}$ is the $\phi$-component of the external magnetic field $B_{\mathrm{ex}}$, which acts to confine the torus plasma in the cylindrical coordinate system. We assume $B_{\mathrm{ex}}$ to be a potential magnetic field. To compute $n_{r}$ at any point, we first select two footpoints of one magnetic field line which has $T_{w} > 0.80$ in the MFR, following our definition of an MFR. We define $X'$- and $Y'$-axes in the horizontal plane, as shown in Figure \ref{fig:cylindrical} (a). The footpoints F1 and F2 were on the $Y'$-axis, and the $X'$-axis was orthogonal to the $Y'$-axis. Second, we define the cylindrical coordinate system whose origin is located at the central point $O'$. Then we can define the axial-radial decay index to evaluate the rate of decrease with distance of the confining magnetic field, in this coordinate system, for any arbitrary direction from the origin ($O'$).

Figure \ref{fig:cylindrical} (b) shows a schematic drawing of this cylindrical coordinate system with the origin located at the center of the MFR. F1 and F2 indicate the two footpoints of the model MFR, and $O'$ is their mid-point, which is defined as the central point of the MFR whose magnetic topology we assumed to be an ideal torus plasma, having an axial electric current flowing inside the torus. The $Z$-axis is the vertical axis with respect to the photosphere. Defining the inclined two axes on the photosphere, we can define a cylindrical coordinate system which should contain a cross section of the MFR on the $X'-Z$ plane. Although the MFR moved slightly with time prior to the eruption, we fix not only the newly-defined $X'$- and $Y'$- axes but also the cross section, the $X'-Z$ plane. We confirmed that the distribution of the magnetic twist number on the fixed cross section in the new coordinate system did not change dramatically during both the formation and the eruption phases of the MFR evolution.

\begin{figure}[htbp!]
\centering    
\includegraphics[width=1.00\textwidth]{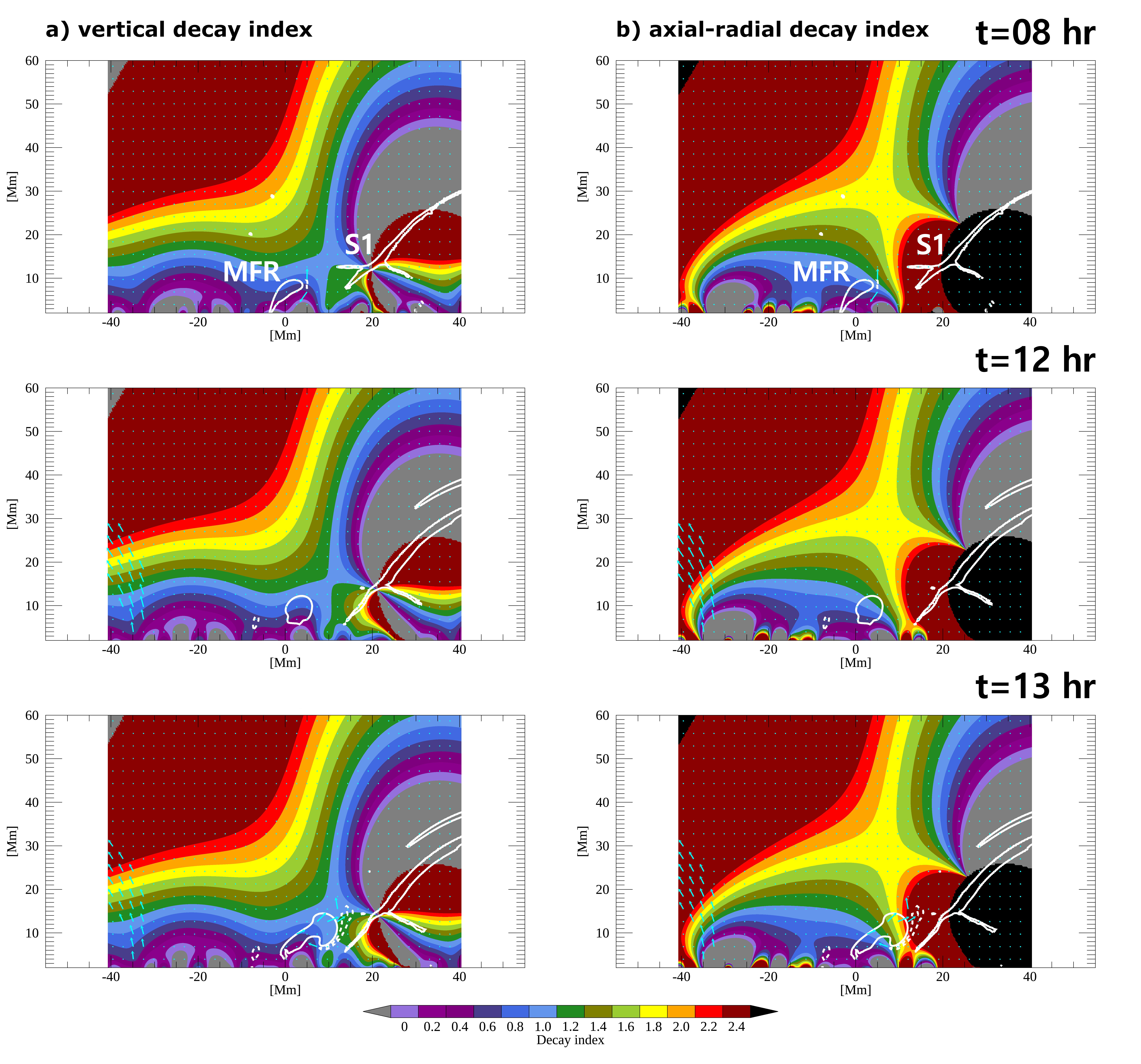}
\caption{Temporal evolution of the two decay indexes, the vertical decay index $n$ (column a) and the axial-radial decay index $n_{r}$ (column b) for the failed eruption. Shown are $n$ and $n_{r}$ on the $X'-Z$ plane cross section as defined in Figure \ref{fig:cylindrical}, at t=8, 12 and 13 hr. The horizontal axis is the $X'$-axis, and the vertical axis is the $Z$-axis. The origin corresponds to the center ($O'$) of the MFR. The $X'-Z$ plane is constructed by the magnetic topology at t=12 hr and fixed afterwards. In the background images in both columns, the black masked areas correspond to the regions where $B_{\phi}<0$, that is, the Lorentz force due to the external field has the opposite direction to the regions where $B_{\phi}>0$.  The gray masked areas indicate regions where $n_{r}<0$, that is, the external field outside was greater than the inside. The white solid and dashed line contours represent the magnetic twist numbers ($T_{w}$) of 0.80 and -0.80 on this plane, respectively. S1 may have a large $T_{w}$ because the connectivity of its associated magnetic field lines changes dramatically, resulting in highly sheared field lines. The cyan arrows show the velocity field on this plane, indicating a velocities between 20 km/s and 80 km/s.}
\label{fig:failed}
\end{figure}

Do these different formulations of the decay index provide different information on the eruptions? To answer this, we investigated the temporal evolution of both the vertical and axial-radial decay indexes ($n$ and $n_{r}$, respectively) over periods covering both the failed and full eruptions. In calculating $n$, we only used the $X'-$ component of $B_{\mathrm{ex}}$. This approach is suitable in this case because we define and set the orientation of the MFR ($O'$), which allows us to compare those two parameters only using the confining field on the $X'-Z$ plane. First the case of the failed eruption: Figure \ref{fig:failed} (a) and Figure \ref{fig:failed} (b) show respectively the temporal evolution of $n$ and $n_{r}$ distributions both in the context of the locations of the MFR and of S1, displayed as a cross section of the QSL-related field lines (see Figure \ref{fig:qsl} (a) for reference).  In both panels, we see that S1 was located in the west side of the MFR, and the MFR moves towards S1. To determine the direction of eruption, we assume that the high velocity field near the MFR's apex describes its eruptive motion. The direction of the eruption with respect to S1 is reasonable because the MFR can be more likely to erupt toward a region where $n_{r}$ achieves $n_{r} \approx 1.5$, however, the eruption was confined ultimately because the MFR could not reach S1. In terms of the distributions of $n$ and $n_{r}$, the distribution of $n_{r}$ is more consistent with the direction of the eruption than the distribution of $n$ because $n < 1.0$ along the direction of propagation, but $n_{r}$ exceeded the critical decay index even achieving $n_{r} \geq 2.0$ along the direction of the eruption. However, here, $n_{r}$ does not provide sufficient information about the boundary effect (or the line tying effect), suggesting that more studies are needed to investigate $n_{r}$ as a discriminator of the eruptivity of MFRs.

\begin{figure}[htbp!] 
\centering    
\includegraphics[width=1.00\textwidth]{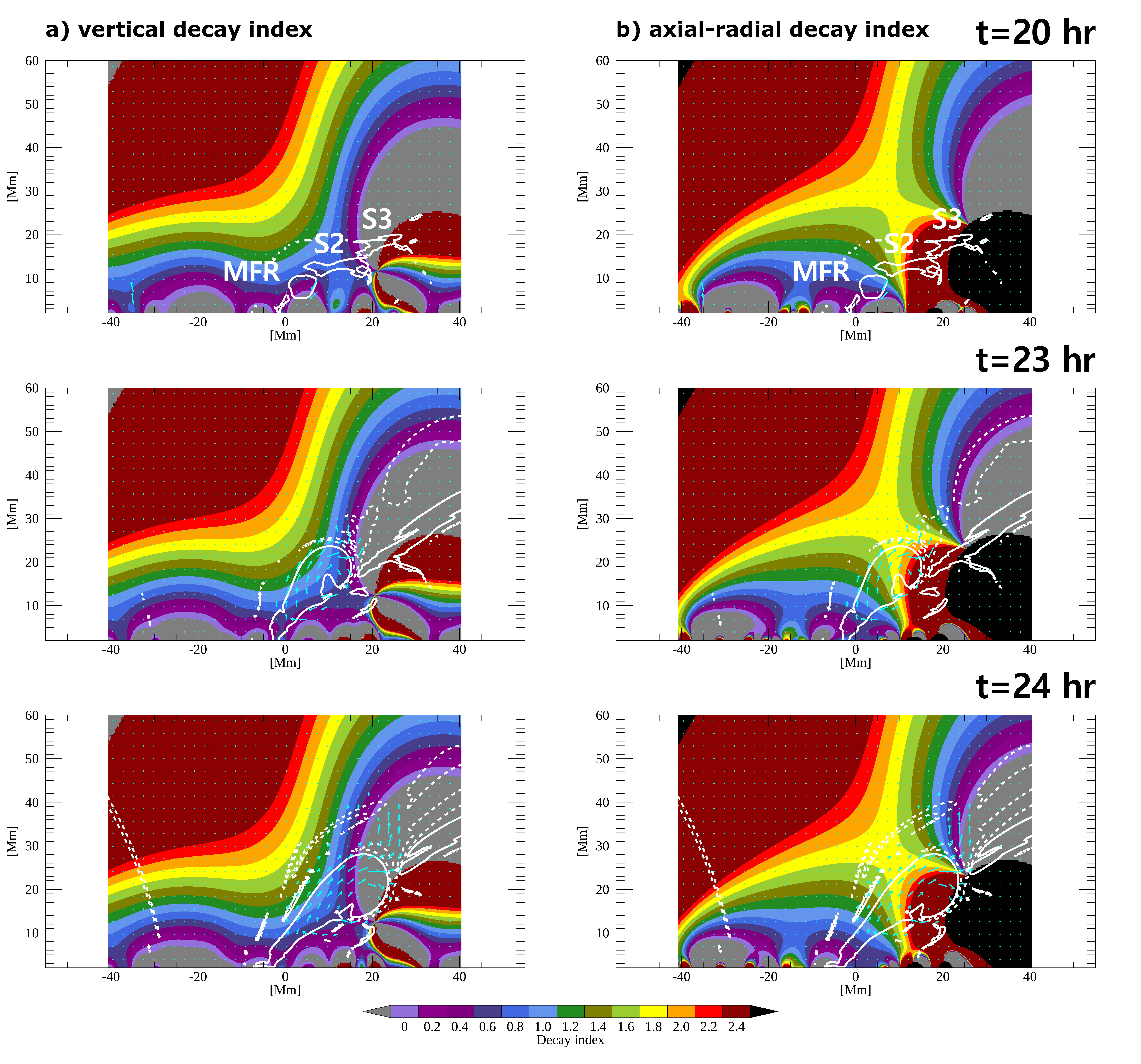}
\caption{Temporal evolution of $n$ and $n_{r}$ indices following Figure \ref{fig:failed} (including the plane shown), here for the full eruption associated with the M5.3 flare, at t=20, 23 and 24 hr. The volumetric distributions of two decay indices did not change dramatically compared to the failed-eruption, and we find that the eruptive direction of the MFR is also similar in this case. However, as shown for t=20 hr, S2 and S3 were present to the west of the MFR.  As the eruption progresses, at t=23 and 24 hr, the MFR merged S2 and S3 successively, and the erupting MFR caused magnetic reconnection with the neighboring field lines at the front of the MFR.}
\label{fig:successful}
\end{figure}

Second, the case of the full eruption: Figure \ref{fig:successful} (a) and (b) show the temporal evolution of $n$ and $n_{r}$, respectively. At t=20 hr, the new QSLs related to S2 and S3 were formed, and the MFR initiated an eruption toward S2 and S3 which merged S2 and S3 during this process. The erupting MFR also caused magnetic reconnection along its propagation path. Similar to the case of the failed eruption, the distribution of $n_{r}$ is more consistent with the direction of the eruption than is the distribution of $n$ because $n_{r}>2.0$ which is larger than the critical decay index, whereas $n$ does not achieve this magnitude.  Unlike the failed eruption, however, in this case the MFR could erupt fully because it reached S2, and magnetic reconnection was enabled in its vicinity.  In other words, the existence of the S2-related QSL likely allowed the MFR to become torus-unstable more easily, and the spatial relationship between S2 and regions of $n_{r} > n_{\rm crit}$ suggests that the torus instability mechanism is one candidate for the initiation of this solar eruption. Furthermore, the path of eruption can be influenced by the decay of the overlying field and its spatial relation to the MFR prior to the eruption.

\subsection{Eruption-model implications} \label{subsec:eruption_mechanism}

%With regard to the eruption mechanism associated with the M5.3 flare, o
Figure \ref{fig:tw1600} shows that the location of the observed flare ribbon was located along the PIL. Our modeling indicates that the simulated magnetic twist flux density ($\tau$) was also situated in those regions. The similarity in these locations suggests that the flare ribbons can be formed through the flare reconnections within the erupting MFR. This differs from the model proposed by \citet{2016NatCo...711522J} which suggested that the flare reconnections occurred in a region distinct from the MFR. The proposed full eruption process as per our data-driven model is as follows: As described in Section \ref{sec:results}, the simulation reproduced that the MFR erupted by the torus instability, and the field lines within the erupting MFR reconnected with the neighboring field lines. Figure \ref{fig:successful} presents evidence that these neighboring field lines exhibited positive helicity, as does the MFR. Thus, reconnection between these magnetic systems can further increase the helicity in the MFR. The reconnected field lines, identified as those having $T_{w}=-0.80$, are also indicated in Figure \ref{fig:successful}. Meanwhile, flare-related reconnection is believed to enable particles to heat the chromosphere near the footpoints of the reconnected field lines (\citealt{1995ApJ...451L..83S}; \citealt{2011LRSP....8....6S}). Thus, in our simulation, the particles in the MFR could be driven toward the footpoints of the MFR as a result of the flare-related reconnection. Our suggestion finds support through the comparison with the observation of the flare ribbons produced by the M5.3 flare because one piece of evidence for the particle movement during a solar flare is the presence of the flare ribbons observed in the chromosphere. The location of the footpoints in our modeling aligned with the position of the flare ribbons, which were situated along the PIL formed by P1 and N1 as shown in Figure \ref{fig:tw1600}. This result clearly demonstrates that the main reconnection driving the M5.3 flare occurred above the PIL between P1 and N1.

Comparing our results to the previous study (\citealt{2016NatCo...711522J}), they reproduced the filament ejection related to the M5.3 flare, suggesting that the reconnection area was located on the north-west side of the MFR.  Our study differs from \citealt{2016NatCo...711522J} primarily in the FOV used, with ours using a smaller FOV for input. That is, the area of negative-polarity magnetic field located to the north-west of the MFR was included in their simulation, and their modeling suggested this additional magnetic flux played an important role in the  filament ejection.  In our modeling, however, the QSL-related field lines L2 and L6 in Figure \ref{fig:qsl} (a) and (b) were formed by N1 and the simulation boundary. Taking into account the distribution of strong vertical magnetic fields on N3, this magnetic topology may be valid despite the limitation of the domain size. The validation of this simulated magnetic topology is also supported by the implication that the field lines extending from the simulation boundary in the west side may exhibit a similar topology to the real coronal fields, as the local field lines are likely to converge toward N3 from all directions, including from the west side.  Nevertheless, this limitation should also yield the different results from the previous study.

 \section{summary} \label{sec:summary}

We have produced a model of a solar eruption related to the eruptive M5.3 flare in AR 11283, using $\vctr{E} \times \vctr{B}$-driven modeling and a limited field of view computational box. In our modeling, we found that there was an accumulation of free magnetic energy in the corona prior to the M5.3 flare.

The M5.3 flare was associated with an inclined eruption of a magnetic flux rope located above the polarity inversion line. To explain the resulting inclined eruption, we demonstrate that an axial-radial decay index ($n_{r}$) in which the magnetic decay index is calculated for all axial-radial directions with respect to the MFR, provides evidence that the magnetic topology provides a path for eruption that is not evident when the model is analyzed using the normal radial decay index. The methodology behind the axial-radial decay index ($n_{r}$) can be additionally used to evaluate the MFR against the growth of the torus instability.  While we find that the axial-radial decay index can explain the inclination of the full eruption, we still cannot uniquely predict the success or failure of MFR eruptions even with this new implementation.  

\section*{acknowledgements}
This research is supported by ``Computational Joint Research Program (Collaborative Research Project on Computer Science with High-Performance Computing)'' at the Institute for Space-Earth Environment Research, Nagoya University and JSPS KAKENHI Grant No. JP21H04492 (PI: K. Kusano).

\clearpage

\bibliography{ms}{}
\bibliographystyle{aasjournal}

\end{document}